\documentclass[11pt]{article}
\usepackage{amsfonts}
\usepackage{amsmath}
\usepackage{amssymb}
\usepackage{indentfirst}
\usepackage{graphicx}

\begin{document}

\title{In\"{o}n\"{u}-Wigner Contraction and $D=2+1$ Supergravity}
\author{P. K. Concha$^{1,2}$\thanks{%
patillusion@gmail.com}, O. Fierro$^{3}$\thanks{%
ofierro@ucsc.cl}, E. K. Rodr\'{\i}guez$^{1,2}$\thanks{%
everodriguezd@gmail.com}, \\
{\small $^{1}$\textit{Departamento de Ciencias, Facultad de Artes Liberales,}%
}\\
{\small \textit{Universidad Adolfo Ib\'{a}\~{n}ez,}}\\
{\small Av. Padre Hurtado 750, Vi\~{n}a del Mar, Chile}\\
{\small $^{2}$\textit{Instituto de Ciencias F\'{\i}sicas y Matem\'{a}ticas,
Universidad Austral de Chile,}}\\
{\small Casilla 567, Valdivia, Chile}\\
$^{3}${\small \textit{Departamento de Matem\'{a}tica y F\'{\i}sica Aplicadas,%
}}\\
\ {\small \textit{Universidad Cat\'{o}lica de la Sant\'{\i}sima Concepci\'{o}%
n,}}\\
\ {\small Alonso de Rivera 2850, Concepci\'{o}n, Chile}}
\maketitle

\begin{abstract}
We present a generalization of the standard In\"{o}n\"{u}-Wigner contraction
by rescaling not only the generators of a Lie superalgebra but also the
arbitrary constants appearing in the components of the invariant tensor. The
procedure presented here allows to obtain explicitly the Chern-Simons
supergravity action of a contracted superalgebra. In particular we show that
the Poincar\'{e} limit can be performed to a $D=2+1$ $\left( p,q\right) $ $%
AdS$ Chern-Simons supergravity in presence of the exotic form. We also
construct a new three-dimensional $\left( 2,0\right) $ Maxwell Chern-Simons
supergravity theory as a particular limit of $\left( 2,0\right) $ $AdS$%
-Lorentz supergravity theory. The generalization for $\mathcal{N}=p+q$
gravitini is also considered.
\end{abstract}

\vspace{-13.5cm}

\begin{flushright}
{\footnotesize UAI-PHY-16/12}
\end{flushright}

\vspace{12cm}

\section{Introduction}

The three-dimensional (super)gravity theory represents an interesting toy
model in order to approach higher dimensional (super)gravity theories, which
are not only more difficult but also leads to tedious calculations.
Additionally, the $D=2+1$ model has the remarkable property to be written as
a gauge theory using the Chern-Simons (CS) formalism \cite{AT, W}. In
particular, the three-dimensional supersymmetric extension of General
Relativity \cite{DK, D} can be obtained as a CS gravity theory using $(A)dS$
or Poincar\'{e} supergroup. A wide class of $\mathcal{N}$-extended
Supergravities and further extensions have been studied in diverse contexts
in e.g., \cite{PvN1, RPvN, AT2, NG, HIPT, BTZ, GTW, GS, ABRHST, BHRT,
BKPRYZ, ABRS, BKNTM, NT, FISV, ABBOS, FMT, CFRS, FMT2, BRZ, BDMT}.

The derivation of a supergravity action for a given superalgebra is not, in
general, a trivial task and its construction is not always assured. \ On the
other hand, several (super)algebras can be obtained as an In\"{o}n\"{u}%
-Wigner (IW) contraction of a given (super)algebra \cite{IW, WW}.
Nevertheless, the Chern-Simons action based on the IW contracted
(super)algebra cannot always be obtained by rescaling the gauge fields and
considering some limit as in the (anti)commutation relations. In particular
it is known that, in presence of the exotic Lagrangian, the Poincar\'{e}
limit cannot be applied to a $\left( p,q\right) $ $AdS$ CS supergravity \cite%
{AT2}. This difficulty can be overcome extending the $\mathfrak{osp}\left(
2,p\right) \otimes \mathfrak{osp}\left( 2,q\right) $ superalgebra by
introducing the automorphism generators $\mathfrak{so}\left( p\right) $ and $%
\mathfrak{so}\left( q\right) $ \cite{HIPT}. In such a case, the IW
contraction can be applied and reproduces the Poincar\'{e} limit leading to
a new $\left( p,q\right) $ Poincar\'{e} supergravity which includes
additional $\mathfrak{so}\left( p\right) \oplus \mathfrak{so}\left( q\right)
$ gauge fields.

Here, we present a generalization of the IW contraction by considering not
only the rescaling of the generators but also the constants of the
non-vanishing components of an invariant tensor. \ The method introduced
here assures the construction of any CS action based on a contracted
(super)algebra. In particular, we show that the Poincar\'{e} limit can be
applied to a $\left( p,q\right) $ $AdS$ supergravity in presence of the
exotic Lagrangian without introducing extra fields as in Ref.~\cite{HIPT}.
Subsequently, we apply the method to different $\left( p,q\right) $ $AdS$%
-Lorentz supergravities whose IW contraction leads to diverse $\left(
p,q\right) $ Maxwell supergravities. The possibility to turn the IW
contraction into an algebraic operation is not new and has already been
presented in the context of asymptotic symmetries and higher spin theories
in Ref.~\cite{KRR}. Other interesting results using diverse flat limit
contractions in supergravity can be found in Ref.~\cite{LM}.

At the bosonic level, the Maxwell symmetries have lead to interesting
gravity theories allowing to recover General Relativity from Chern-Simons
and Born-Infeld (BI) theories \cite{GRCS, CPRS1, CPRS2, CPRS3}. On the other
hand, the $AdS$-Lorentz and its generalizations allow to recover the Pure
Lovelock \cite{Cai:2006pq, Dadhich:2012ma, Dadhich:2015ivt} Lagrangian in a
matter-free configuration from CS and BI theories \cite{CDIMR, CMR}. At the
supersymmetric level, the Maxwell superalgebra provides a pure supergravity
action in the MacDowell-Mansouri formalism \cite{CR2}. More recently, a
three-dimensional CS action based on the minimal Maxwell superalgebra has
been presented in Ref.~\cite{CFRS} using the expansion procedure. Here, we
show that the same result can be obtained using our alternative approach.
Besides, we show that the Maxwell limit can also be applied in a $\left(
p,q\right) $ enlarged supergravity leading to a $\left( p,q\right) $ Maxwell
supergravity with an exotic Lagrangian.

The organization of the present work is as follows: In section 2, we apply
our approach to $\mathcal{N}=1$ and $\mathcal{N}=p+q$ $AdS$ CS
supergravities. In particular, we show that the Poincar\'{e} limit can be
applied to $\left( p,q\right) $ $AdS$ CS supergravity theories in presence
of the exotic Lagrangian. In section 3, we discuss the In\"{o}n\"{u}-Wigner
contraction of an expanded supergravity. In particular, we describe the
general scheme. In section 4, we apply our procedure to a $\mathcal{N}=1$
expanded CS supergravity. In section 5, we present the CS formulation of the
$\left( 2,0\right) $ and $\left( p,q\right) $ Maxwell supergravities and
discuss their relations to $\left( 2,0\right) $ and $\left( p,q\right) $ $%
AdS $-Lorentz supergravities, respectively. Section 6 concludes our work
with some comments and possible developments.

\section{In\"{o}n\"{u}-Wigner contraction and the invariant tensor}

The standard In\"{o}n\"{u}-Wigner contraction \cite{IW, WW} of a Lie
(super)algebra $\mathfrak{g}$ consists basically in properly rescaling the
generators by a parameter $\sigma $ and applying the limit $\sigma
\rightarrow \infty $ corresponding to a contracted (super)algebra.

Despite having the proper contracted (super)algebra following the IW scheme,
the contracted invariant tensor cannot be trivially obtained. This is
particularly regrettable since the invariant tensor is an essential
ingredient in the construction of a Chern-Simons action.

In this paper, we present a generalization of the\ standard In\"{o}n\"{u}%
-Wigner contraction considering the rescaling not only of the generators but
also of the constants appearing in the invariant tensor. \ The method
introduced here allows to obtain the non-vanishing components of the
invariant tensor of an IW contracted (super)algebra. Thus, the construction
of any CS action based on an IW contracted (super)algebra is assured. In
particular, we apply the method to different $\left( p,q\right) $ $AdS$%
-Lorentz superalgebras whose IW contraction leads to diverse $\left(
p,q\right) $ Maxwell superalgebras.

Let us first apply the approach to the $AdS\ $supergravity in order to
derive the Poincar\'{e} supergravity.

\subsection{Poincar\'{e}$\,\ $and $\mathfrak{osp}\left( 2|1\right) \otimes
\mathfrak{sp}\left( 2\right) \ $supergravity}

As in the bosonic level, the IW contraction of the $AdS$ superalgebra leads
to the Poincar\'{e} one. Besides, the Poincar\'{e} CS supergravity action
can be obtained considering a particular limit after an appropriate
rescaling of the fields of the super $AdS$ CS action. Nevertheless, in
presence of torsion the exotic Lagrangian, which has no Poincar\'{e} limit
\cite{AT2}, is added to the $AdS$ CS supergravity.

The three-dimensional Chern-Simons action is given by%
\begin{equation}
I_{CS}^{\left( 2+1\right) }=k\int \left\langle AdA+\frac{2}{3}%
A^{3}\right\rangle \,,  \label{CS}
\end{equation}%
where $A$ corresponds to the gauge connection one-form and $\left\langle {%
\dots }\right\rangle $ denotes the invariant tensor. In the case of the $%
\mathfrak{osp}\left( 2|1\right) \otimes \mathfrak{sp}\left( 2\right) \ $%
superalgebra, the connection one-form is given by%
\begin{equation}
A=\frac{1}{2}\omega ^{ab}\tilde{J}_{ab}+\frac{1}{l}e^{a}\tilde{P}_{a}+\frac{1%
}{\sqrt{l}}\psi ^{\alpha }\tilde{Q}_{\alpha }\,,
\end{equation}%
where $\tilde{J}_{ab}$, $\tilde{P}_{a}$ and $\tilde{Q}_{\alpha }$ are the $%
\mathfrak{osp}\left( 2|1\right) \otimes \mathfrak{sp}\left( 2\right) $
generators. The gauge fields $e^{a}$, $\omega ^{ab}$ and $\psi $ are the
dreibein, the spin connection and the gravitino, respectively. Here, the
length scale $l$ is introduced purposely in order to have dimensionless
generators $T_{A}=\left\{ \tilde{J}_{ab},\tilde{P}_{a},\tilde{Q}_{\alpha
}\right\} $ such that the connection one-form $A=A_{\mu }^{A}T_{A}dx^{\mu }$
must also be dimensionless. Since the dreibein $e^{a}=e_{\mu }^{a}dx^{\mu }$
is related to the spacetime metric $g_{\mu \nu }$ through $g_{\mu \nu
}=e_{\mu }^{a}e_{\nu }^{b}\eta _{ab}$, it must have dimensions of length.
Then, the "true" gauge field should be considered as $e^{a}/l$. In the same
way, we consider $\psi /\sqrt{l}$ as the supersymmetry gauge field since the
gravitino $\psi =\psi _{\mu }dx^{\mu }$ has dimensions of (length)$^{1/2}$.

The (anti)-commutation relations for the $\mathfrak{osp}\left( 2|1\right)
\otimes \mathfrak{sp}\left( 2\right) $ superalgebra are given by
\begin{eqnarray}
\left[ \tilde{J}_{ab},\tilde{J}_{cd}\right] &=&\eta _{bc}\tilde{J}_{ad}-\eta
_{ac}\tilde{J}_{bd}-\eta _{bd}\tilde{J}_{ac}+\eta _{ad}\tilde{J}_{bc}\,, \\
\left[ \tilde{J}_{ab},\tilde{P}_{c}\right] &=&\eta _{bc}\tilde{P}_{a}-\eta
_{ac}\tilde{P}_{b}\,, \\
\left[ \tilde{P}_{a},\tilde{P}_{b}\right] &=&\tilde{J}_{ab}\,, \\
\left[ \tilde{J}_{ab},\tilde{Q}_{\alpha }\right] &=&-\frac{1}{2}\left(
\Gamma _{ab}\tilde{Q}\right) _{\alpha }\,,\text{ \ \ \ }\left[ \tilde{P}_{a},%
\tilde{Q}_{\alpha }\right] =-\frac{1}{2}\left( \Gamma _{a}\tilde{Q}\right)
_{\alpha }\,, \\
\left\{ \tilde{Q}_{\alpha },\tilde{Q}_{\beta }\right\} &=&-\frac{1}{2}\left[
\left( \Gamma ^{ab}C\right) _{\alpha \beta }\tilde{J}_{ab}-2\left( \Gamma
^{a}C\right) _{\alpha \beta }\tilde{P}_{a}\right] \,,
\end{eqnarray}%
where $C$ denotes the charge conjugation matrix, $\Gamma _{\alpha }$
represents the Dirac matrices and $\Gamma _{ab}=\frac{1}{2}\left[ \Gamma
_{a},\Gamma _{b}\right] $.

The non-vanishing components of an invariant tensor for the $\mathfrak{osp}%
\left( 2|1\right) \otimes \mathfrak{sp}\left( 2\right) $ superalgebra are
given by%
\begin{eqnarray}
\left\langle \tilde{J}_{ab}\tilde{P}_{c}\right\rangle &=&\mu _{1}\epsilon
_{abc}\,,  \label{AdSIT1} \\
\left\langle \tilde{J}_{ab}\tilde{J}_{cd}\right\rangle &=&\mu _{0}\left(
\eta _{ad}\eta _{bc}-\eta _{ac}\eta _{bd}\right) \,, \\
\left\langle \tilde{P}_{a}\tilde{P}_{b}\right\rangle &=&\mu _{0}\eta _{ab}\,,
\\
\left\langle \tilde{Q}_{\alpha }\tilde{Q}_{\beta }\right\rangle &=&2\left(
\mu _{1}-\mu _{0}\right) C_{\alpha \beta }\,,  \label{AdSIT4}
\end{eqnarray}%
where $\mu _{0}$ and $\mu _{1}$ are arbitrary constants. Then, considering
the invariant tensor (\ref{AdSIT1})-(\ref{AdSIT4}) and the connection
one-form in the general expression for the $D=3$ CS action we have%
\begin{eqnarray}
I_{CS}^{\left( 2+1\right) } &=&k\int \left[ \frac{\mu _{0}}{2}\left( \omega
_{\text{ }b}^{a}d\omega _{\text{ }a}^{b}+\frac{2}{3}\omega _{\text{ }%
c}^{a}\omega _{\text{ }b}^{c}\omega _{\text{ }a}^{b}+\frac{2}{l^{2}}%
e^{a}T_{a}-\frac{2}{l}\bar{\psi}\Psi \right) \right.  \notag \\
&&\left. +\frac{\mu _{1}}{l}\left( \epsilon _{abc}R^{ab}e^{c}+\frac{1}{3l^{2}%
}\epsilon _{abc}e^{a}e^{b}e^{c}+2\bar{\psi}\Psi \right) -d\left( \frac{\mu
_{1}}{2l}\epsilon _{abc}\omega ^{ab}e^{c}\right) \right] \,,
\end{eqnarray}%
where%
\begin{eqnarray*}
R^{ab} &=&d\omega ^{ab}+\omega _{\text{ }c}^{a}\omega ^{cb}\,,\text{ \ \ }%
T^{a}=de^{a}+\omega _{\text{ }b}^{a}e^{b}\,, \\
\Psi &=&d\psi +\frac{1}{4}\omega _{ab}\Gamma ^{ab}\psi +\frac{1}{2l}%
e_{a}\Gamma ^{a}\psi \,.
\end{eqnarray*}

It is known that the following rescaling of the generators%
\begin{equation*}
\tilde{J}_{ab}\rightarrow \bar{J}_{ab},\text{ \ \ \ }\tilde{P}%
_{a}\rightarrow \sigma ^{2}\bar{P}_{a},\text{ \ \ \ \ }\tilde{Q}_{\alpha
}\rightarrow \sigma \bar{Q}_{\alpha }
\end{equation*}%
leads to the Poincar\'{e} superalgebra in the limit $\sigma \rightarrow
\infty $. \ \ It seems natural to construct a Poincar\'{e} CS supergravity
action combining the corresponding rescaling of the generators with the $AdS$
invariant tensor given by Eqs. (\ref{AdSIT1})-(\ref{AdSIT4}). However, such
rescaling of the generators leads to a trivial invariant tensor and then to
a trivial CS action. In order to obtain the right Poincar\'{e} limit at the
level of the action, a rescaling of the arbitrary constants appearing in the
invariant tensor should also be considered. \ Indeed, a rescaling which
preserves the curvatures structure is given by%
\begin{equation*}
\mu _{0}\rightarrow \mu _{0}\,,\text{ \ \ \ \ }\mu _{1}\rightarrow \sigma
^{2}\mu _{1}.
\end{equation*}%
Then, considering the rescaling of both the generators and the constants,
one can see that the limit $\sigma \rightarrow \infty $ leads to the
non-vanishing components of the invariant tensor for the Poincar\'{e}
superalgebra,%
\begin{eqnarray}
\left\langle \bar{J}_{ab}\bar{P}_{c}\right\rangle _{\mathcal{P}} &=&\mu
_{1}\epsilon _{abc}\,, \\
\left\langle \bar{J}_{ab}\bar{J}_{cd}\right\rangle _{\mathcal{P}} &=&\mu
_{0}\left( \eta _{ad}\eta _{bc}-\eta _{ac}\eta _{bd}\right) \,, \\
\left\langle \bar{Q}_{\alpha }\bar{Q}_{\beta }\right\rangle _{\mathcal{P}}
&=&2\mu _{1}C_{\alpha \beta }\,,
\end{eqnarray}%
where $\bar{J}_{ab}$, $\bar{P}_{a}$ and $\bar{Q}_{\alpha }$ are the Poincar%
\'{e} generators. Considering the Poincar\'{e} gauge connection one-form and
the non-vanishing components of the Poincar\'{e} invariant tensor, the CS
action reduces to%
\begin{equation}
I_{CS}^{\left( 2+1\right) }=k\int \frac{\mu _{0}}{2}\left( \omega _{\text{ }%
b}^{a}d\omega _{\text{ }a}^{b}+\frac{2}{3}\omega _{\text{ }c}^{a}\omega _{%
\text{ }b}^{c}\omega _{\text{ }a}^{b}\right) +\frac{\mu _{1}}{l}\left(
\epsilon _{abc}R^{ab}e^{c}+2\bar{\psi}\Psi \right) -d\left( \frac{\mu _{1}}{%
2l}\epsilon _{abc}\omega ^{ab}e^{c}\right) \,,
\end{equation}%
where the fermionic curvature is now given by%
\begin{equation*}
\Psi =d\psi +\frac{1}{4}\omega _{ab}\Gamma ^{ab}\psi \,.
\end{equation*}%
Let us note that the present approach allows to trivially obtain the Poincar%
\'{e} limit from the $\mathfrak{osp}\left( 2|1\right) \otimes \mathfrak{sp}%
\left( 2\right) $ CS action. One could suggest that the same result can be
obtained considering $l\rightarrow \infty \,\,$, nevertheless the presence
of the exotic Lagrangian forbids such limit. \ Additionally, one can notice
that the gravitino does not contribute anymore to the exotic form.

\subsection{$\left( p,q\right) $ Poincar\'{e} and $\mathfrak{osp}\left(
2|p\right) \otimes \mathfrak{osp}\left( 2|q\right) $ supergravity}

Let us now consider the $\left( p,q\right) $ $AdS$ supergravity theories,
which can be viewed as a direct sum of $AdS$ superalgebras. \ It is well
known that the $\left( p,q\right) $ Poincar\'{e} superalgebra can be derived
as a In\"{o}n\"{u}-Wigner contraction of the $\left( p,q\right) $ $AdS$
superalgebra. However, as was mentioned in Refs.~\cite{HIPT, GTW, AI}, the
Poincar\'{e} limit at the level of the action requires enlargement of the $%
AdS$ superalgebra, considering a direct sum of the $\mathfrak{so}\left(
p\right) \oplus \mathfrak{so}\left( q\right) $ algebra and the $\left(
p,q\right) $ $AdS$ superalgebra. Here, we show that our approach allows to
obtain the Poincar\'{e} limit without introducing additional gauge fields.
In particular, the non-vanishing components of the invariant tensor of the $%
\left( p,q\right) $ Poincar\'{e} superalgebra are obtained from the $AdS$
ones.

The supersymmetric extension of the $AdS$ algebra contains $\mathcal{N}=p+q$
gravitinos, and it is spanned by the set of generators $\left\{ \text{%
\thinspace \thinspace }\tilde{J}_{ab},\tilde{P}_{a},\tilde{T}^{ij},\tilde{T}%
^{IJ},\tilde{Q}_{\alpha }^{i},Q_{\alpha }^{I}\right\} $ which satisfy \cite%
{HIPT}%
\begin{eqnarray}
\left[ \tilde{J}_{ab},\tilde{J}_{cd}\right] &=&\eta _{bc}\tilde{J}_{ad}-\eta
_{ac}\tilde{J}_{bd}-\eta _{bd}\tilde{J}_{ac}+\eta _{ad}\tilde{J}_{bc}\,, \\
\left[ \tilde{T}^{ij},\tilde{T}^{kl}\right] &=&\delta ^{jk}\tilde{T}%
^{il}-\delta ^{ik}\tilde{T}^{jl}-\delta ^{jl}\tilde{T}^{ik}+\delta ^{il}%
\tilde{T}^{jk}\text{\thinspace }, \\
\left[ \tilde{T}^{IJ},\tilde{T}^{KL}\right] &=&\delta ^{JK}\tilde{T}%
^{IL}-\delta ^{IK}\tilde{T}^{JL}-\delta ^{JL}\tilde{T}^{IK}+\delta ^{IL}%
\tilde{T}^{JK}\text{\thinspace }, \\
\left[ \tilde{J}_{ab},\tilde{P}_{c}\right] &=&\eta _{bc}\tilde{P}_{a}-\eta
_{ac}\tilde{P}_{b}\,,\text{ \ \ \ \ }\left[ \tilde{P}_{a},\tilde{P}_{b}%
\right] =\tilde{J}_{ab}\,, \\
\left[ \tilde{T}^{ij},\tilde{Q}_{\alpha }^{k}\right] &=&\left( \delta ^{jk}%
\tilde{Q}_{\alpha }^{i}-\delta ^{ik}\tilde{Q}_{\alpha }^{j}\right) \,,\text{
\ \ }\left[ \tilde{T}^{IJ},\tilde{Q}_{\alpha }^{K}\right] =\left( \delta
^{JK}\tilde{Q}_{\alpha }^{I}-\delta ^{IK}\tilde{Q}_{\alpha }^{J}\right) \,,
\\
\left[ \tilde{J}_{ab},\tilde{Q}_{\alpha }^{i}\right] &=&-\frac{1}{2}\left(
\Gamma _{ab}\tilde{Q}^{i}\right) _{\alpha }\,,\text{ \ \ \ }\left[ \tilde{P}%
_{a},\tilde{Q}_{\alpha }^{i}\right] =-\frac{1}{2}\left( \Gamma _{a}\tilde{Q}%
^{i}\right) _{\alpha }\,, \\
\left[ \tilde{J}_{ab},\tilde{Q}_{\alpha }^{I}\right] &=&-\frac{1}{2}\left(
\Gamma _{ab}\tilde{Q}^{I}\right) _{\alpha }\,,\text{ \ \ \ }\left[ \tilde{P}%
_{a},\tilde{Q}_{\alpha }^{I}\right] =\frac{1}{2}\left( \Gamma _{a}\tilde{Q}%
^{I}\right) _{\alpha }\,, \\
\left\{ \tilde{Q}_{\alpha }^{i},\tilde{Q}_{\beta }^{j}\right\} &=&-\frac{1}{2%
}\delta ^{ij}\left[ \left( \Gamma ^{ab}C\right) _{\alpha \beta }\tilde{J}%
_{ab}-2\left( \Gamma ^{a}C\right) _{\alpha \beta }\tilde{P}_{a}\right]
+C_{\alpha \beta }\tilde{T}^{ij}\,, \\
\left\{ \tilde{Q}_{\alpha }^{I},\tilde{Q}_{\beta }^{J}\right\} &=&\frac{1}{2}%
\delta ^{IJ}\left[ \left( \Gamma ^{ab}C\right) _{\alpha \beta }\tilde{J}%
_{ab}+2\left( \Gamma ^{a}C\right) _{\alpha \beta }\tilde{P}_{a}\right]
-C_{\alpha \beta }\tilde{T}^{IJ}\,,
\end{eqnarray}%
where $i,j=1,\dots ,p$ and $I,J=1,\dots ,q$. Here, the $\tilde{T}^{ij}$ and $%
\tilde{T}^{IJ}$ generators correspond to internal symmetry generators and
satisfy a $\mathfrak{so}\left( p\right) $ and $\mathfrak{so}\left( q\right) $
algebra, respectively.

One can introduce the $\mathfrak{osp}\left( 2|p\right) \times \mathfrak{osp}%
\left( 2|q\right) $ connection one-form $A$ given by%
\begin{equation}
A=\frac{1}{2}\omega ^{ab}\tilde{J}_{ab}+\frac{1}{l}e^{a}\tilde{P}_{a}+\frac{1%
}{2}A^{ij}\tilde{T}_{ij}+\frac{1}{2}A^{IJ}\tilde{T}_{IJ}+\frac{1}{\sqrt{l}}%
\bar{\psi}_{i}\tilde{Q}^{i}+\frac{1}{\sqrt{l}}\bar{\psi}_{I}\tilde{Q}^{I}\,.
\end{equation}

The non-vanishing components of the invariant tensor for the $(p,q)$ $AdS$
superalgebra are given by%
\begin{eqnarray}
\left\langle \tilde{J}_{ab}\tilde{J}_{cd}\right\rangle &=&\mu _{0}\left(
\eta _{ad}\eta _{bc}-\eta _{ac}\eta _{bd}\right) \,, \\
\left\langle \tilde{J}_{ab}\tilde{P}_{c}\right\rangle &=&\mu _{1}\epsilon
_{abc}\,, \\
\left\langle \tilde{P}_{a}\tilde{P}_{b}\right\rangle &=&\mu _{0}\eta _{ab}\,,
\\
\left\langle \tilde{Q}_{\alpha }^{i}\tilde{Q}_{\beta }^{j}\right\rangle
&=&2\left( \mu _{1}-\mu _{0}\right) C_{\alpha \beta }\delta ^{ij}\,, \\
\left\langle \tilde{Q}_{\alpha }^{I}\tilde{Q}_{\beta }^{J}\right\rangle
&=&2\left( \mu _{1}+\mu _{0}\right) C_{\alpha \beta }\delta ^{IJ}\,, \\
\left\langle \tilde{T}^{ij}\tilde{T}^{kl}\right\rangle &=&2\left( \mu
_{0}-\mu _{1}\right) \left( \delta ^{il}\delta ^{kj}-\delta ^{ik}\delta
^{lj}\right) \,, \\
\left\langle \tilde{T}^{IJ}\tilde{T}^{KL}\right\rangle &=&2\left( \mu
_{0}+\mu _{1}\right) \left( \delta ^{IL}\delta ^{KJ}-\delta ^{IK}\delta
^{LJ}\right) \,,
\end{eqnarray}%
where $\mu _{0}$ and $\mu _{1}$ are arbitrary constants. Considering the
connection one-form $A$ and the non-vanishing components of the invariant
tensor in the three-dimensional CS general expression (\ref{CS}), we obtain
the $\mathfrak{osp}\left( 2|p\right) \otimes \mathfrak{osp}\left( 2|q\right)
$ supergravity action in three dimensions:
\begin{eqnarray}
I_{CS}^{\left( 2+1\right) } &=&k\int \frac{\mu _{0}}{2}\left( \omega _{\text{
}b}^{a}d\omega _{\text{ }a}^{b}+\frac{2}{3}\omega _{\text{ }c}^{a}\omega _{%
\text{ }b}^{c}\omega _{\text{ }a}^{b}+\frac{2}{l^{2}}e^{a}T_{a}\right)
\notag \\
&&+\frac{\mu _{1}}{l}\epsilon _{abc}\left( R^{ab}e^{c}+\frac{1}{3l^{2}}%
e^{a}e^{b}e^{c}\right)  \notag \\
&&+\left( \mu _{0}-\mu _{1}\right) \left[ A^{ij}dA^{ji}+\frac{2}{3}%
A^{ik}A^{kj}A^{ji}\right]  \notag \\
&&+\left( \mu _{0}+\mu _{1}\right) \left[ A^{IJ}dA^{JI}+\frac{2}{3}%
A^{IK}A^{KJ}A^{JI}\right]  \notag \\
&&+2\left( \mu _{1}-\mu _{0}\right) \left[ \frac{1}{l}\bar{\psi}^{i}\left(
\Psi ^{i}+A^{ij}\psi ^{j}\right) \right] \,  \notag \\
&&+2\left( \mu _{1}+\mu _{0}\right) \,\left[ \frac{1}{l}\bar{\psi}^{I}\left(
\Psi ^{I}+A^{IJ}\psi ^{J}\right) \right] \,,
\end{eqnarray}%
where%
\begin{eqnarray*}
R^{ab} &=&d\omega ^{ab}+\omega _{\text{ }c}^{a}\omega ^{cb}\,,\text{ \ \ \ \
}T^{a}=de^{a}+\omega _{\text{ }c}^{a}e^{c}\,, \\
\Psi ^{i} &=&d\psi ^{i}+\frac{1}{4}\omega _{ab}\Gamma ^{ab}\psi ^{i}+\frac{1%
}{2l}e_{a}\Gamma ^{a}\psi ^{i}\,, \\
\Psi ^{I} &=&d\psi ^{I}+\frac{1}{4}\omega _{ab}\Gamma ^{ab}\psi ^{I}-\frac{1%
}{2l}e_{a}\Gamma ^{a}\psi ^{I}\,.
\end{eqnarray*}%
Let us note that an off-shell formulation for $\mathfrak{osp}\left(
2|p\right) \times \mathfrak{osp}\left( 2|q\right) $ supergravity is not
assured when $p+q>1$. Interestingly, diverse off-shell formulations for $%
\left( p,q\right) $ $AdS$ supergravity when $p+q\leq 3$ can be found in
Refs.~\cite{KTM, KLTM}.

One can note that the $\left( p,q\right) $ Poincar\'{e} superalgebra can be
obtained from the $\left( p,q\right) $ $AdS$ one considering the following
rescaling of the generators%
\begin{equation*}
\tilde{J}_{ab}\rightarrow \bar{J}_{ab},\text{ \ \ \ }\tilde{T}%
^{ij}\rightarrow \sigma ^{2}\bar{T}^{ij},\text{ \ \ \ \ }\tilde{P}%
_{a}\rightarrow \sigma ^{2}\bar{P}_{a},\text{ \ \ \ \ }\tilde{Q}_{\alpha
}\rightarrow \sigma \bar{Q}_{\alpha }
\end{equation*}%
and the limit $\sigma \rightarrow \infty $:%
\begin{eqnarray}
\left[ \bar{J}_{ab},\bar{J}_{cd}\right] &=&\eta _{bc}\bar{J}_{ad}-\eta _{ac}%
\bar{J}_{bd}-\eta _{bd}\bar{J}_{ac}+\eta _{ad}\bar{J}_{bc}\,, \\
\left[ \bar{J}_{ab},\bar{P}_{c}\right] &=&\eta _{bc}\bar{P}_{a}-\eta _{ac}%
\bar{P}_{b}\,, \\
\left[ \bar{J}_{ab},\bar{Q}_{\alpha }^{i}\right] &=&-\frac{1}{2}\left(
\Gamma _{ab}\bar{Q}^{i}\right) _{\alpha }\,,\text{ } \\
\left[ \bar{J}_{ab},\bar{Q}_{\alpha }^{I}\right] &=&-\frac{1}{2}\left(
\Gamma _{ab}\bar{Q}^{I}\right) _{\alpha }\,,\text{ \ } \\
\left\{ \bar{Q}_{\alpha }^{i},\bar{Q}_{\beta }^{j}\right\} &=&\delta
^{ij}\left( \Gamma ^{a}C\right) _{\alpha \beta }\bar{P}_{a}+C_{\alpha \beta }%
\bar{T}^{ij}\,, \\
\left\{ \bar{Q}_{\alpha }^{I},\bar{Q}_{\beta }^{J}\right\} &=&\delta
^{IJ}\left( \Gamma ^{a}C\right) _{\alpha \beta }\bar{P}_{a}-C_{\alpha \beta }%
\bar{T}^{IJ}\,,
\end{eqnarray}%
Here, $T^{ij}$ and $T^{IJ}$ behave as central charges and no longer satisfy
a $\mathfrak{so}\left( p\right) $ and a $\mathfrak{so}\left( q\right) $
algebra, respectively. Indeed, when $p$ or $q$ is greater than $1$, the $%
\left( p,q\right) $ Poincar\'{e} superalgebra corresponds to a central
extension of the $\mathcal{N}$-extended Poincar\'{e} superalgebra.

At the level of the action, we have to consider a rescaling of the constants
appearing in the invariant tensor. \ Indeed, a rescaling which preserves the
curvatures structure is given by%
\begin{equation*}
\mu _{0}\rightarrow \mu _{0}\,,\text{ \ \ \ \ }\mu _{1}\rightarrow \sigma
^{2}\mu _{1}.
\end{equation*}%
Then, the limit $\sigma \rightarrow \infty $ leads to the non-vanishing
components of the invariant tensor for the Poincar\'{e} superalgebra,%
\begin{eqnarray}
\left\langle \bar{J}_{ab}\bar{J}_{cd}\right\rangle _{\mathcal{P}} &=&\mu
_{0}\left( \eta _{ad}\eta _{bc}-\eta _{ac}\eta _{bd}\right) \,,  \label{PIT1}
\\
\left\langle \bar{J}_{ab}\bar{P}_{c}\right\rangle _{\mathcal{P}} &=&\mu
_{1}\epsilon _{abc}\,, \\
\left\langle \bar{Q}_{\alpha }^{i}\bar{Q}_{\beta }^{j}\right\rangle _{%
\mathcal{P}} &=&2\mu _{1}C_{\alpha \beta }\delta ^{ij}\,,  \label{PIT3} \\
\left\langle \bar{Q}_{\alpha }^{I}\bar{Q}_{\beta }^{J}\right\rangle _{%
\mathcal{P}} &=&2\mu _{1}C_{\alpha \beta }\delta ^{IJ}\,,  \label{PIT4}
\end{eqnarray}%
where $\bar{J}_{ab},\bar{P}_{a},\bar{Q}_{\alpha }^{i}$ and $\bar{Q}_{\alpha
}^{I}$ correspond now to the Poincar\'{e} generators. As was noticed in Ref.~%
\cite{HIPT}, there are no components of the $\left( p,q\right) $ Poincar\'{e}
invariant tensor including the $\bar{T}^{ij}$ and $\bar{T}^{IJ}$ generators.
Indeed, considering the $\left( p,q\right) $ Poincar\'{e} connection
one-form and the invariant tensor (\ref{PIT1})-(\ref{PIT4}) in the general
expression for the CS action we find%
\begin{eqnarray}
I_{CS}^{\left( 2+1\right) } &=&k\int \frac{\mu _{0}}{2}\left( \omega _{\text{
}b}^{a}d\omega _{\text{ }a}^{b}+\frac{2}{3}\omega _{\text{ }c}^{a}\omega _{%
\text{ }b}^{c}\omega _{\text{ }a}^{b}\right)  \notag \\
&&+\frac{\mu _{1}}{l}\left( \epsilon _{abc}R^{ab}e^{c}+2\bar{\psi}^{i}\Psi
^{i}+2\bar{\psi}^{I}\Psi ^{I}\right) \,,  \label{pqCS}
\end{eqnarray}%
where%
\begin{eqnarray*}
\Psi ^{i} &=&d\psi ^{i}+\frac{1}{4}\omega _{ab}\Gamma ^{ab}\psi ^{i}, \\
\Psi ^{I} &=&d\psi ^{I}+\frac{1}{4}\omega _{ab}\Gamma ^{ab}\psi ^{I}\,.
\end{eqnarray*}%
The Poincar\'{e} CS action (\ref{pqCS}) can be directly obtained from the $%
\left( p,q\right) $ $AdS$ one considering the rescaling of the constants ($%
\mu _{0}\rightarrow \mu _{0}\,,$ \ \ \ \ $\mu _{1}\rightarrow \sigma ^{2}\mu
_{1}$), the rescaling of the fields:%
\begin{equation*}
\omega _{ab}\rightarrow \omega _{ab},\text{ \ \ \ }A^{ij}\rightarrow \sigma
^{-2}A^{ij},\text{ \ \ \ \ }e_{a}\rightarrow \sigma ^{-2}e_{a},\text{ \ \ \
\ }\psi ^{i}\rightarrow \sigma ^{-1}\psi ^{i},\text{ \ \ \ \ }\psi
^{I}\rightarrow \sigma ^{-1}\psi ^{I}\,,
\end{equation*}%
and the limit $\sigma \rightarrow \infty $. Thus, the Poincar\'{e} limit can
be applied without introducing extra fields and/or change of basis.
Nevertheless, as in the $\mathcal{N}=1$ case, the gravitino does not
contribute to the exotic form. Besides, no $\mathfrak{so}\left( p\right) $
or $\mathfrak{so}\left( q\right) $ gauge fields appear in the Lagrangian. In
order to obtain more interesting supergravity actions whose gravitino
appears in the exotic term, it is necessary to consider our approach to
enlarged supersymmetries.

On the other hand, let us note that the term proportional to $\mu _{1}$
reproduces the action of Ref.~\cite{HIPT} when $F(A)=0$. Naturally, the same
procedure can be applied to the direct sum of $\left( p,q\right) $ $AdS$
superalgebra and $\mathfrak{so}\left( p\right) \oplus \mathfrak{so}\left(
q\right) $ algebra, which would lead to the most general action for $\left(
p,q\right) $ Poincar\'{e} superalgebra \cite{HIPT}.

\section{In\"{o}n\"{u}-Wigner contraction of an $S$-expanded supergravity}

The development of the Lie (super)algebra expansion method has played an
important role in order to derive new (super)gravity theories \cite{HS,
AIPV, AIPV2, AIPV3, Sexp, Caroca2010a, Caroca2010b, Caroca2011, CKMN, AMNT,
ACCSP, Durka, ILPR}. In particular, the semigroup expansion method ($S$%
-expansion) allows to find explicitly the non-vanishing components of an
invariant tensor for an expanded (super)algebra in terms of the original one
\cite{Sexp}. This feature is particularly useful since the invariant tensor
is a crucial ingredient in the construction of a Chern-Simons action.

Nevertheless, the CS action for a contracted superalgebra cannot be naively
obtained by rescaling the generators appearing in the non-vanishing
components of an invariant tensor and considering some limit.

As in the previous section, we show that by applying the rescaling to both
generators and constants of the invariant tensor, the CS action for a
contracted superalgebra can be obtained. In particular, we present the
general scheme in order to derive an IW contracted supergravity from an $S$%
-expanded one.

First, we shall consider the contraction of the following subspace
decomposition of the Lie $S$-expanded superalgebra $\mathfrak{G}=S\times
\mathfrak{g}$,%
\begin{equation}
\mathfrak{G}=W_{0}\oplus W_{1}\oplus W_{2}\text{\thinspace },
\end{equation}%
where $W_{0}$ corresponds to a subalgebra, $W_{1}$ corresponds to the
fermionic subspace and $W_{2}$ is generated by boost generators. Such
decomposition satisfies%
\begin{eqnarray}
\left[ W_{0},W_{0}\right] &\subset &W_{0}\,,\text{ \ \ \ \ \ \ \ \ \ \ \ \ }%
\left[ W_{0},W_{2}\right] \subset W_{2}\,, \\
\left[ W_{0},W_{1}\right] &\subset &W_{1}\,,\text{ \ \ \ \ \ \ \ \ \ \ \ \ }%
\left[ W_{1},W_{2}\right] \subset W_{1}\,, \\
\left[ W_{1},W_{1}\right] &\subset &W_{0}\oplus W_{2}\text{\thinspace },%
\text{ \ \ \ \ \ }\left[ W_{2},W_{2}\right] \subset W_{0}\,,
\end{eqnarray}%
where each subspace is generated by sets of generators%
\begin{equation*}
W_{p}=\left\{ X_{p}^{\left( i\right) }=\lambda _{i+p}X_{p}\text{ with }%
i=0,2,4\dots ,n-p\text{ and }p=0,1,2\right\} \,.
\end{equation*}%
Here $X_{p}$ are the generators of the original superalgebra $\mathfrak{g}$
and $\lambda _{i+p}$ is an element of a semigroup $S$ satisfying some
explicit multiplication law of the $S_{\mathcal{M}}$ family \cite{SS}.

The IW contraction of $\mathfrak{G}$ is obtained considering the rescaling
of the expanded generators%
\begin{equation*}
X_{p}^{\left( i\right) }=\sigma ^{i+p}X_{p}^{\left( i\right) }
\end{equation*}%
and applying the limit $\sigma \rightarrow \infty $.

On the other hand, according to Theorem VII.2 of Ref.~\cite{Sexp}, the
invariant tensor for an $S$-expanded superalgebra $\mathfrak{G}$ can be
obtained from the original ones through%
\begin{equation}
\left\langle T_{\left( A,j\right) }T_{\left( B,k\right) }\right\rangle _{%
\mathfrak{G}}=\tilde{\alpha}_{i}K_{jk}^{\text{ \ \ }i}\left\langle
T_{A}T_{B}\right\rangle _{\mathfrak{g}}\,,
\end{equation}%
with $T_{\left( A,j\right) }=\lambda _{j}T_{A}$. Here $\tilde{\alpha}_{i}$
are arbitrary constants and $K_{jk}^{\text{ \ \ }i}$ is the $2$-selector for
the semigroup $S$ defined as%
\begin{equation*}
K_{jk}^{\text{ \ \ }i}=\left\{
\begin{array}{c}
1,\text{ when }i=i\left( j,k\right) \\
0,\text{ otherwise,\ \ \ \ \ \ \ \ }%
\end{array}%
\right.
\end{equation*}%
with $\lambda _{i\left( j,k\right) }=\lambda _{j}\lambda _{k}$.

The IW contraction of the invariant tensor is obtained, considering the
rescaling of the generators%
\begin{equation*}
T_{\left( A,j\right) }=\sigma ^{j}T_{\left( A,j\right) },
\end{equation*}%
the rescaling of the constant $\tilde{\alpha}_{i}$,%
\begin{equation*}
\tilde{\alpha}_{i}\rightarrow \sigma ^{i}\tilde{\alpha}_{i}
\end{equation*}%
and applying the limit $\sigma \rightarrow \infty $.

The approach considered here offers a close relation between expansion and
contraction. Interestingly, as we shall see, our procedure allows us to
obtain the contracted supergravity in presence of expanded
Pontryagin-Chern-Simons form.

In the following sections, we shall present known and new Maxwell
supergravities considering the present IW approach to diverse $S$-expanded
supergravities.

\section{In\"{o}n\"{u}-Wigner contraction and $\mathcal{N}=1$ supergravity}

Here, we present a generalization of the In\"{o}n\"{u}-Wigner contraction
combining the $S$-expansion method and the rescaling of the invariant tensor
and generators. In particular, we show the $D=3$ CS action based on a $%
\mathcal{N}=1$ Maxwell superalgebra.

A non-standard supersymmetrization of the Maxwell algebra was introduced in
Refs.~\cite{Sorokas1, Sorokas2} which can be obtained as an IW contraction
of the standard $AdS$-Lorentz superalgebra%
\footnote{Also known as Poincar\'{e}
semi-simple extended superalgebra.}
\cite{Sorokas3, DKGS}. Nevertheless, the non-standard Maxwell supersymmetric
action and its physical relevance remains poorly explored due to its unusual
anticommutation relations. Indeed, the $P_{a}$ generators of the
non-standard Maxwell superalgebra are not expressed as bilinear expressions
of the fermionic generators $Q$,%
\begin{equation*}
\left\{ Q_{\alpha },Q_{\beta }\right\} =-\frac{1}{2}\left( \Gamma
^{ab}C\right) _{\alpha \beta }Z_{ab}\,.
\end{equation*}%
This feature prevents construction of a supergravity action based on this
peculiar supersymmetry. \ Despite this particularity, there is an
alternative in order to construct a $\mathcal{N}=1$ supergravity action
based on the Maxwell supersymmetries.

A particular Maxwell superalgebra, also called as the minimal
supersymmetrization of the Maxwell algebra \cite{BGKL1, BGKL2, L, KL},
differs from the non-standard Maxwell one since it possesses an additional
fermionic generator. Interestingly, a minimal Maxwell superalgebra can be
derived as an In\"{o}n\"{u}-Wigner contraction of a new minimal $AdS$%
-Lorentz superalgebra introduced in Ref.~\cite{CRS}.

Before studying the explicit IW contraction at the level of the invariant
tensor, we first present the explicit construction of a CS supergravity
invariant under the minimal $AdS$-Lorentz superalgebra. To this purpose, we
will apply the $S$-expansion procedure analogously to the four-dimensional
case \cite{CRS}.

\subsection{Minimal $AdS$-Lorentz exotic supergravity}

Following the procedure of Ref.~\cite{CRS}\thinspace , a minimal \ $AdS$%
-Lorentz superalgebra can be derived as an $S$-expansion of the $\mathfrak{%
osp}\left( 2|1\right) \otimes \mathfrak{sp}\left( 2\right) $ superalgebra. \
Indeed, considering $S_{\mathcal{M}}^{\left( 4\right) }=\left\{ \lambda
_{0},\lambda _{1},\lambda _{2},\lambda _{3},\lambda _{4}\right\} $ as the
abelian semigroup whose elements satisfy%
\begin{equation}
\lambda _{\alpha }\lambda _{\beta }=\left\{
\begin{array}{c}
\lambda _{\alpha +\beta },\text{ \ }\ \ \ \text{\ if }\alpha +\beta \leq 4%
\text{ \ \ \ } \\
\lambda _{\alpha +\beta -4},\text{ \ \ if }\alpha +\beta >4\text{\ \ \ \ }%
\end{array}%
\right.
\end{equation}%
and after extracting a resonant subalgebra of $S_{\mathcal{M}}^{\left(
4\right) }\times \left( \mathfrak{osp}\left( 2|1\right) \otimes \mathfrak{sp}%
\left( 2\right) \right) $, the minimal AdS-Lorentz superalgebra is obtained
\cite{CRS}. This algebra corresponds to a supersymmetric extension of the $%
\mathfrak{so}\left( 2,2\right) \oplus \mathfrak{so}\left( 2,1\right) $
algebra $=\left\{ J_{ab},P_{a},Z_{ab}\right\} $ and is generated by $\left\{
J_{ab},P_{a},\tilde{Z}_{ab},\tilde{Z}_{a},Z_{ab},Q_{\alpha },\Sigma _{\alpha
}\right\} $. This superalgebra, as in the Maxwell case, is quite different
from the standard AdS-Lorentz superalgebra discussed in Refs.~\cite{FISV,
Sorokas3}. In fact, besides extra bosonic generators $\left\{ \tilde{Z}_{ab},%
\tilde{Z}_{a}\right\} $, it also has more than one spinor generator. The
explicit (anti)commutation relations can be found in Appendix A for $%
\mathcal{N}=1$.

The construction of a Chern-Simons action for the minimal $AdS$-Lorentz
superalgebra requires the gauge connection one-form $A$:%
\begin{equation}
A=\frac{1}{2}\omega ^{ab}J_{ab}+\frac{1}{2}\tilde{k}^{ab}\tilde{Z}_{ab}+%
\frac{1}{2}k^{ab}Z_{ab}+\frac{1}{l}e^{a}P_{a}+\frac{1}{l}\tilde{h}^{a}\tilde{%
Z}_{a}+\frac{1}{\sqrt{l}}\psi ^{\alpha }Q_{\alpha }+\frac{1}{\sqrt{l}}\xi
^{\alpha }\Sigma _{\alpha }\,.  \label{1FC3}
\end{equation}%
Since we have considered a dimensionless connection one-form, a factor $l$
has to be introduced for the dreibein field and the dreibein like field $%
\tilde{h}^{a}$. \ The same argument applies for the spinor fields.

Another crucial ingredient necessary to write down a CS supergravity action
is the invariant tensor. Following the Theorem VII.2 of Ref.~\cite{Sexp},
the non-vanishing components of an invariant tensor for\ the super minimal $%
AdS$-Lorentz are given by%
\begin{eqnarray}
\left\langle J_{ab}J_{cd}\right\rangle _{\mathcal{S}} &=&\tilde{\alpha}%
_{0}\left\langle \tilde{J}_{ab}\tilde{J}_{cd}\right\rangle =\alpha
_{0}\left( \eta _{ad}\eta _{bc}-\eta _{ac}\eta _{bd}\right) \,,
\label{invt3a} \\
\left\langle J_{ab}\tilde{Z}_{cd}\right\rangle _{\mathcal{S}} &=&\tilde{%
\alpha}_{2}\left\langle \tilde{J}_{ab}\tilde{J}_{cd}\right\rangle =\alpha
_{2}\left( \eta _{ad}\eta _{bc}-\eta _{ac}\eta _{bd}\right) \,, \\
\left\langle \tilde{Z}_{ab}Z_{cd}\right\rangle _{\mathcal{S}} &=&\tilde{%
\alpha}_{2}\left\langle \tilde{J}_{ab}\tilde{J}_{cd}\right\rangle =\alpha
_{2}\left( \eta _{ad}\eta _{bc}-\eta _{ac}\eta _{bd}\right) \,,
\end{eqnarray}%
\begin{eqnarray}
\left\langle J_{ab}Z_{cd}\right\rangle _{\mathcal{S}} &=&\tilde{\alpha}%
_{4}\left\langle \tilde{J}_{ab}\tilde{J}_{cd}\right\rangle =\alpha
_{4}\left( \eta _{ad}\eta _{bc}-\eta _{ac}\eta _{bd}\right) \,, \\
\left\langle \tilde{Z}_{ab}\tilde{Z}_{cd}\right\rangle _{\mathcal{S}} &=&%
\tilde{\alpha}_{4}\left\langle \tilde{J}_{ab}\tilde{J}_{cd}\right\rangle
=\alpha _{4}\left( \eta _{ad}\eta _{bc}-\eta _{ac}\eta _{bd}\right) \,\,, \\
\left\langle Z_{ab}Z_{cd}\right\rangle _{\mathcal{S}} &=&\tilde{\alpha}%
_{4}\left\langle \tilde{J}_{ab}\tilde{J}_{cd}\right\rangle =\alpha
_{4}\left( \eta _{ad}\eta _{bc}-\eta _{ac}\eta _{bd}\right) \,\,,
\end{eqnarray}%
\begin{eqnarray}
\left\langle J_{ab}P_{c}\right\rangle _{\mathcal{S}} &=&\left\langle
Z_{ab}P_{c}\right\rangle _{\mathcal{S}}=\left\langle \tilde{Z}_{ab}\tilde{Z}%
_{c}\right\rangle _{\mathcal{S}}=\tilde{\alpha}_{2}\left\langle \tilde{J}%
_{ab}\tilde{P}_{c}\right\rangle =\beta _{2}\epsilon _{abc}\,, \\
\left\langle J_{ab}\tilde{Z}_{c}\right\rangle _{\mathcal{S}} &=&\left\langle
Z_{ab}\tilde{Z}_{c}\right\rangle _{\mathcal{S}}=\left\langle \tilde{Z}%
_{ab}P_{c}\right\rangle _{\mathcal{S}}=\tilde{\alpha}_{4}\left\langle \tilde{%
J}_{ab}\tilde{P}_{c}\right\rangle =\beta _{4}\epsilon _{abc}\,, \\
\left\langle P_{a}P_{b}\right\rangle _{\mathcal{S}} &=&\left\langle \tilde{Z}%
_{a}\tilde{Z}_{b}\right\rangle _{\mathcal{S}}=\tilde{\alpha}_{4}\left\langle
\tilde{P}_{a}\tilde{P}_{b}\right\rangle =\alpha _{4}\eta _{ab}\,, \\
\left\langle P_{a}\tilde{Z}_{b}\right\rangle _{\mathcal{S}} &=&\tilde{\alpha}%
_{2}\left\langle \tilde{P}_{a}\tilde{P}_{b}\right\rangle =\alpha _{2}\eta
_{ab}\,,  \label{invt3c} \\
\left\langle Q_{\alpha }Q_{\beta }\right\rangle _{\mathcal{S}}
&=&\left\langle \Sigma _{\alpha }\Sigma _{\beta }\right\rangle _{\mathcal{S}%
}=\tilde{\alpha}_{2}\left\langle \tilde{Q}_{\alpha }\tilde{Q}_{\beta
}\right\rangle =2\left( \beta _{2}-\alpha _{2}\right) C_{\alpha \beta }\,, \\
\left\langle Q_{\alpha }\Sigma _{\beta }\right\rangle _{\mathcal{S}} &=&%
\tilde{\alpha}_{4}\left\langle \tilde{Q}_{\alpha }\tilde{Q}_{\beta
}\right\rangle =2\left( \beta _{4}-\alpha _{4}\right) C_{\alpha \beta }\,,
\label{invt3b}
\end{eqnarray}%
where $\left\{ \tilde{J}_{ab},\tilde{P}_{a},\tilde{Q}_{\alpha }\right\} $
generate the $\mathfrak{osp}\left( 2|1\right) \otimes \mathfrak{sp}\left(
2\right) $ superalgebra (see eqs. (\ref{AdSIT1})-(\ref{AdSIT4})) and where
we have defined%
\begin{align*}
\alpha _{0}& \equiv \tilde{\alpha}_{0}\mu _{0}\,,\text{ \ \ \ \ }\alpha
_{2}\equiv \tilde{\alpha}_{2}\mu _{0}\,,\text{ \ \ \ \ }\alpha _{4}\equiv
\tilde{\alpha}_{4}\mu _{0}\,, \\
\beta _{2}& \equiv \tilde{\alpha}_{2}\mu _{1}\,,\text{ \ \ \ \ }\beta
_{4}\equiv \tilde{\alpha}_{4}\mu _{1}\,.\text{\ }
\end{align*}%
Here $\tilde{\alpha}_{0},\tilde{\alpha}_{2},\tilde{\alpha}_{4}$ are
arbitrary constants as $\mu _{0}$ and $\mu _{1}$. The CS supergravity action
can be written considering the connection one-form (\ref{1FC3}) and the
non-vanishing component of the invariant (\ref{invt3a})-(\ref{invt3b}) in
the general three-dimensional CS expression%
\begin{equation*}
I_{CS}^{\left( 2+1\right) }=k\int \left\langle AdA+\frac{2}{3}%
A^{3}\right\rangle \,.
\end{equation*}%
Thus, we have modulo boundary terms%
\begin{eqnarray}
I_{CS}^{\left( 2+1\right) } &=&k\int \frac{\alpha _{0}}{2}\left[ \omega _{%
\text{ }b}^{a}d\omega _{\text{ }a}^{b}+\frac{2}{3}\omega _{\text{ }%
c}^{a}\omega _{\text{ }b}^{c}\omega _{\text{ }a}^{b}\right] +\frac{\beta _{2}%
}{l}\left[ \epsilon _{abc}\left( R^{ab}e^{c}+\frac{1}{3l^{2}}%
e^{a}e^{b}e^{c}+K^{ab}e^{c}\right. \right.  \notag \\
&&\left. \left. \tilde{K}^{ab}\tilde{h}^{c}+\frac{1}{l^{2}}\tilde{h}^{a}%
\tilde{h}^{b}e\right) +2\bar{\psi}\Psi +2\bar{\xi}\Xi \right]  \notag \\
&&+\alpha _{2}\left[ R_{\text{ }b}^{a}\tilde{k}_{\text{ }a}^{b}+K_{\text{ }%
b}^{a}\tilde{k}_{\text{ }a}^{b}+\tilde{K}_{\text{ }b}^{a}k_{\text{ }a}^{b}+%
\frac{1}{l^{2}}e^{a}H_{a}+\frac{1}{l^{2}}\tilde{h}^{a}K_{a}-\frac{2}{l}\bar{%
\psi}\Psi -\frac{2}{l}\bar{\xi}\Xi \right]  \notag \\
&&+\frac{\beta _{4}}{l}\left[ \epsilon _{abc}\left( R^{ab}\tilde{h}^{c}+%
\frac{1}{3l^{2}}\tilde{h}^{a}\tilde{h}^{b}\tilde{h}^{c}+\tilde{K}%
^{ab}e^{c}+K^{ab}\tilde{h}^{c}+\frac{1}{l^{2}}e^{a}e^{b}\tilde{h}^{c}\right)
+2\bar{\xi}\Psi +2\bar{\psi}\Xi \right]  \notag \\
&&+\alpha _{4}\left[ R_{\text{ }b}^{a}k_{\text{ }a}^{b}+K_{\text{ }b}^{a}k_{%
\text{ }a}^{b}+\tilde{K}_{\text{ }b}^{a}\tilde{k}_{\text{ }a}^{b}+\frac{1}{%
l^{2}}e^{a}K_{a}+\frac{1}{l^{2}}\tilde{h}^{a}H_{a}-\frac{2}{l}\bar{\xi}\Psi -%
\frac{2}{l}\bar{\psi}\Xi \right] \,,  \label{CSAL}
\end{eqnarray}%
where%
\begin{eqnarray*}
K^{ab} &=&Dk^{ab}+k_{\text{ }d}^{a}k_{\text{ }b}^{d}+\tilde{k}_{\text{ }%
d}^{a}\tilde{k}_{\text{ }b}^{d}\,,\text{ \ \ \ }\tilde{K}^{ab}=D\tilde{k}%
^{ab}+k_{\text{ }d}^{a}\tilde{k}_{\text{ }b}^{d}+k_{b}^{\text{ }d}\tilde{k}_{%
\text{ }d}^{a}\,, \\
H^{a} &=&D\tilde{h}^{a}+k_{\text{ }b}^{a}\tilde{h}^{b}+\tilde{k}_{\text{ }%
b}^{a}e^{b}\,,\text{ \ \ \ \ \ \ }K^{a}=T^{a}+k_{\text{ }b}^{a}e^{b}+\tilde{k%
}_{\text{ }b}^{a}\tilde{h}^{b}\,, \\
\Psi &=&d\psi ^{i}+\frac{1}{4}\omega _{ab}\Gamma ^{ab}\psi ^{i}+\frac{1}{4}%
k_{ab}\Gamma ^{ab}\psi ^{i}+\frac{1}{4}\tilde{k}_{ab}\Gamma ^{ab}\xi ^{i}+%
\frac{1}{2l}e_{a}\Gamma ^{a}\xi ^{i}+\frac{1}{2l}\tilde{h}_{a}\Gamma
^{a}\psi ^{i}\,, \\
\Xi ^{i} &=&d\xi ^{i}+\frac{1}{4}\omega _{ab}\Gamma ^{ab}\xi ^{i}+\frac{1}{4}%
k_{ab}\Gamma ^{ab}\xi ^{i}+\frac{1}{4}\tilde{k}_{ab}\Gamma ^{ab}\psi ^{i}+%
\frac{1}{2l}e_{a}\Gamma ^{a}\psi ^{i}+\frac{1}{2l}\tilde{h}_{a}\Gamma
^{a}\xi ^{i}\,\,.
\end{eqnarray*}%
The CS action (\ref{CSAL}) is locally gauge invariant under the minimal $AdS$%
-Lorentz superalgebra and is split into five independent pieces proportional
to $\alpha _{0},\alpha _{2},\alpha _{4},\beta _{2}$ and $\beta _{4.}$. In
particular, the term proportional to $\alpha _{0}$ corresponds to the exotic
form, while the $\alpha _{2}$ and $\alpha _{4}$ terms contain exotic like
Lagrangians plus fermionic terms.

Let us note that the CS action (\ref{CSAL}) reproduces the three-dimensional
generalized cosmological constant term introduced in Refs.~\cite{SS, AKL}
when $\tilde{k}^{ab}=\tilde{h}^{a}=0$. A generalized supersymmetric
cosmological term has also been introduced in a four-dimensional
MacDowell-Mansouri like action constructed out of the curvature two-form,
based on an $AdS$-Lorentz superalgebra \cite{CRS, CIRR}. Besides, the
bosonic part corresponds to the $AdS$-Lorentz CS action presented in Refs.~%
\cite{DFIMRSV, HR}.

\subsection{The Maxwell limit}

One is tempted to consider an appropriate rescaling of the fields at the
level of the action (\ref{CSAL}) and apply some limit in order to derive the
Maxwell supergravity action. However, although a minimal Maxwell
superalgebra can be obtained as an IW\ contraction of the minimal $AdS$%
-Lorentz superalgebra, the IW contraction of the action (\ref{CSAL}) would
reproduce a trivial CS action. To reproduce a non-trivial CS supergravity
action based on the $\mathcal{N}=1$ Maxwell symmetries, it is necessary to
extend the rescaling of the generators to the\ $\alpha $ and $\beta $
constants appearing in the non-vanishing components of the invariant tensor.
A rescaling which preserves the curvatures structure is given by%
\begin{equation*}
\alpha _{4}\rightarrow \sigma ^{4}\alpha _{4}\,,\text{ \ \ \ \ }\beta
_{4}\rightarrow \sigma ^{4}\beta _{4}\,,\text{ \ \ \ \ }\alpha
_{2}\rightarrow \sigma ^{2}\alpha _{2}\,,\text{ \ \ \ \ }\beta
_{2}\rightarrow \sigma ^{2}\beta _{2}\,,\text{ \ \ \ \ }\alpha
_{0}\rightarrow \alpha _{0}\,.
\end{equation*}%
Then, considering the rescaling of the constants and the generators%
\begin{align*}
\tilde{Z}_{ab}& \rightarrow \sigma ^{2}\tilde{Z}_{ab}\,,\text{ \ \ }%
Z_{ab}\rightarrow \sigma ^{4}Z_{ab}\,,\text{ \ \ }P_{a}\rightarrow \sigma
^{2}P_{a}\,,\text{ \ \ \ \ }J_{ab}\rightarrow J_{ab}\,, \\
\tilde{Z}_{a}& \rightarrow \sigma ^{4}\tilde{Z}_{a}\,,\text{ \ \ }Q_{\alpha
}\rightarrow \sigma Q_{\alpha }\,,\text{\ \ }\Sigma \rightarrow \sigma
^{3}\Sigma \,,
\end{align*}%
and applying the limit $\sigma \rightarrow \infty $, we recover the $%
\mathcal{N}=1$ Maxwell non-vanishing components of the invariant tensor,%
\begin{align}
\left\langle J_{ab}J_{cd}\right\rangle _{\mathcal{M}}& =\alpha _{0}\left(
\eta _{ad}\eta _{bc}-\eta _{ac}\eta _{bd}\right) \,,  \label{inv01} \\
\text{\ \ }\left\langle J_{ab}\tilde{Z}_{cd}\right\rangle _{\mathcal{M}}&
=\alpha _{2}\left( \eta _{ad}\eta _{bc}-\eta _{ac}\eta _{bd}\right) \,, \\
\left\langle \tilde{Z}_{ab}\tilde{Z}_{cd}\right\rangle _{\mathcal{M}}&
=\alpha _{4}\left( \eta _{ad}\eta _{bc}-\eta _{ac}\eta _{bd}\right) \,, \\
\left\langle J_{ab}Z_{cd}\right\rangle _{\mathcal{M}}& =\alpha _{4}\left(
\eta _{ad}\eta _{bc}-\eta _{ac}\eta _{bd}\right) \,,
\end{align}%
\begin{align}
\left\langle J_{ab}P_{c}\right\rangle _{\mathcal{M}}& =\beta _{2}\epsilon
_{abc}\,, \\
\left\langle \tilde{Z}_{ab}P_{c}\right\rangle _{\mathcal{M}}& =\left\langle
J_{ab}\tilde{Z}_{c}\right\rangle _{\mathcal{M}}=\beta _{4}\epsilon _{abc}\,,
\\
\left\langle P_{a}P_{b}\right\rangle _{\mathcal{M}}& =\alpha _{4}\eta
_{ab}\,, \\
\left\langle Q_{\alpha }Q_{\beta }\right\rangle _{\mathcal{M}}& =2\left(
\beta _{2}-\alpha _{2}\right) C_{\alpha \beta }\,, \\
\left\langle Q_{\alpha }\Sigma _{\beta }\right\rangle _{\mathcal{M}}&
=2\left( \beta _{4}-\alpha _{4}\right) C_{\alpha \beta }\,.  \label{inv08}
\end{align}%
Here, the generators now satisfy the (anti)commutation relations of a
minimal Maxwell superalgebra (see Appendix B for $p=1,q=0$). Then using the
Maxwell connection one-form%
\begin{equation*}
A=\frac{1}{2}\omega ^{ab}J_{ab}+\frac{1}{2}\tilde{k}^{ab}\tilde{Z}_{ab}+%
\frac{1}{2}k^{ab}Z_{ab}+\frac{1}{l}e^{a}P_{a}+\frac{1}{l}\tilde{h}^{a}\tilde{%
Z}_{a}+\frac{1}{\sqrt{l}}\psi ^{\alpha }Q_{\alpha }+\frac{1}{\sqrt{l}}\xi
^{\alpha }\Sigma _{\alpha }\,.
\end{equation*}%
and the non-vanishing components of the invariant tensor, \ we derive the
three-dimensional CS supergravity action for the $\mathcal{N}=1$ Maxwell
superalgebra \cite{CFRS}:%
\begin{align}
I_{\mathcal{M}-CS}^{\left( 2+1\right) }& =k\int_{M}\left[ \frac{\alpha _{0}}{%
2}\left( \omega _{\text{ }b}^{a}d\omega _{\text{ }a}^{b}+\frac{2}{3}\omega _{%
\text{ }c}^{a}\omega _{\text{ }b}^{c}\omega _{\text{ }a}^{b}\right) +\frac{%
\beta _{2}}{l}\left( \epsilon _{abc}R^{ab}e^{c}+2\bar{\psi}\Psi \right)
\right.  \notag \\
& +\alpha _{2}\left( R_{\text{ }b}^{a}\tilde{k}_{\text{ }a}^{b}-\frac{2}{l}%
\bar{\psi}\Psi \right) +\frac{\beta _{4}}{l}\left( \epsilon _{abc}\left(
R^{ab}\tilde{h}^{c}+D_{\omega }\tilde{k}^{ab}e^{c}\right) +2\bar{\xi}\Psi +2%
\bar{\psi}\Xi \right)  \notag \\
& \left. +\alpha _{4}\left( R_{\text{ }b}^{a}k_{\text{ }a}^{b}+\frac{1}{l^{2}%
}e^{a}T_{a}+D_{\omega }\tilde{k}_{\text{ }b}^{a}\tilde{k}_{\text{ }a}^{b}-%
\frac{2}{l}\bar{\xi}\Psi -\frac{2}{l}\bar{\psi}\Xi \right) \right] \,,
\end{align}%
where%
\begin{eqnarray*}
R^{ab} &=&d\omega ^{ab}+\omega _{\text{ }c}^{a}\omega ^{cb}\,,\text{ \ \ \ \
}T^{a}=de^{a}+\omega _{\text{ }c}^{a}e^{c}\,, \\
\Psi &=&d\psi +\frac{1}{4}\omega _{ab}\Gamma ^{ab}\psi \,, \\
\Xi &=&d\xi +\frac{1}{4}\omega _{ab}\Gamma ^{ab}\xi +\frac{1}{4}\tilde{k}%
_{ab}\Gamma ^{ab}\psi +\frac{1}{2l}e_{a}\Gamma ^{a}\psi \,.
\end{eqnarray*}%
Thus, we reproduce the CS action presented in Ref.~\cite{CFRS} using a
different approach. Unlike the non-standard Maxwell, the minimal Maxwell
superalgebra allows to properly write a\ three-dimensional CS supergravity
action invariant under local Maxwell supersymmetry transformations (up to
boundary terms). \ However, as was shown at the algebraic level, this
requires the introduction of an additional Majorana spinor field $\xi $. The
presence of a second spinorial generator was already introduced in Refs.~%
\cite{AF, Green} in the context of $D=11$ supergravity and superstring
theory, respectively. Subsequently, the introduction of a second spinorial
generator in the Maxwell symmetries was proposed in Ref.~\cite{BGKL1}. More
recently, a family of Maxwell superalgebras was presented, which generalize
the superalgebra introduced by D 'Auria, Fr\'{e} and Green and contains the
minimal Maxwell one \cite{AILW, CR1, CDMR}.

Let us note that the bosonic term reproduces the CS action presented in
Refs.~\cite{HR, SSV}.

\section{In\"{o}n\"{u}-Wigner contraction and $\mathcal{N}$-extended
Supergravity}

We now present the explicit derivation of the three-dimensional $\left(
p,q\right) $ Maxwell supergravity action using our approach. \ In
particular, we first show the explicit construction of the $\left(
2,0\right) $ $AdS$-Lorentz supergravity using the semigroup expansion
method. The $\left( 2,0\right) $ Maxwell supergravity is then derived from
the $\left( 2,0\right) $ $AdS$-Lorentz superalgebra applying the IW
contraction to both generators and constants appearing in the invariant
tensor.

\subsection{$\mathcal{N}=2$ $AdS$-Lorentz Supergravity}

A non-trivial $\mathcal{N}=2$ $AdS$-Lorentz superalgebra can be obtained as
an $S$-expansion of the $\mathfrak{osp}\left( 2|2\right) \otimes \mathfrak{sp%
}\left( 2\right) $ superalgebra. In order to apply the $S$-expansion
procedure, let us first consider a decomposition of the original superalgebra%
\begin{equation*}
\mathfrak{osp}\left( 2|2\right) \otimes \mathfrak{sp}\left( 2\right)
=V_{0}\oplus V_{1}\oplus V_{2}\,,
\end{equation*}%
where $V_{0}$ corresponds to a subalgebra generated by the Lorentz
generators $\tilde{J}_{ab}$ and by the internal symmetry generator $\tilde{T}%
^{ij}$ (with $i=1,2$), $V_{1}$ corresponds to the fermionic subspace and $%
V_{2}$ is generated by $\tilde{P}_{a}$. In particular, the $\mathfrak{osp}%
\left( 2|2\right) \otimes \mathfrak{sp}\left( 2\right) $ generators satisfy
the following (anti)commutation relations:%
\begin{eqnarray}
\left[ \tilde{J}_{ab},\tilde{J}_{cd}\right] &=&\eta _{bc}\tilde{J}_{ad}-\eta
_{ac}\tilde{J}_{bd}-\eta _{bd}\tilde{J}_{ac}+\eta _{ad}\tilde{J}_{bc}\,, \\
\left[ \tilde{T}^{ij},\tilde{T}^{kl}\right] &=&\delta ^{jk}\tilde{T}%
^{il}-\delta ^{ik}\tilde{T}^{jl}-\delta ^{jl}\tilde{T}^{ik}+\delta ^{il}%
\tilde{T}^{jk}\text{\thinspace }, \\
\left[ \tilde{J}_{ab},\tilde{P}_{c}\right] &=&\eta _{bc}\tilde{P}_{a}-\eta
_{ac}\tilde{P}_{b}\,, \\
\left[ \tilde{P}_{a},\tilde{P}_{b}\right] &=&\tilde{J}_{ab}\,, \\
\left[ \tilde{T}^{ij},\tilde{Q}_{\alpha }^{k}\right] &=&\left( \delta ^{jk}%
\tilde{Q}_{\alpha }^{i}-\delta ^{ik}\tilde{Q}_{\alpha }^{j}\right) \,, \\
\left[ \tilde{J}_{ab},\tilde{Q}_{\alpha }^{i}\right] &=&-\frac{1}{2}\left(
\Gamma _{ab}\tilde{Q}^{i}\right) _{\alpha }\,,\text{ \ \ \ }\left[ \tilde{P}%
_{a},\tilde{Q}_{\alpha }^{i}\right] =-\frac{1}{2}\left( \Gamma _{a}\tilde{Q}%
^{i}\right) _{\alpha }\,, \\
\left\{ \tilde{Q}_{\alpha }^{i},\tilde{Q}_{\beta }^{j}\right\} &=&-\frac{1}{2%
}\delta ^{ij}\left[ \left( \Gamma ^{ab}C\right) _{\alpha \beta }\tilde{J}%
_{ab}-2\left( \Gamma ^{a}C\right) _{\alpha \beta }\tilde{P}_{a}\right]
+C_{\alpha \beta }\tilde{T}^{ij}\,.
\end{eqnarray}%
The subspace structure satisfies%
\begin{eqnarray}
\left[ V_{0},V_{0}\right] &\subset &V_{0}\,,\text{ \ \ \ \ \ \ \ \ \ }\left[
V_{0},V_{1}\right] \subset V_{1}\,,\text{ \ \ \ \ }\left[ V_{0},V_{2}\right]
\subset V_{2}\,,  \label{SD1} \\
\left[ V_{1},V_{1}\right] &\subset &V_{0}\oplus V_{2}\,,\text{ \ \ }\left[
V_{1},V_{2}\right] \subset V_{1}\,,\text{ \ \ \ \ }\left[ V_{2},V_{2}\right]
\subset V_{0}\text{.}  \label{SD2}
\end{eqnarray}%
Let us now consider $S_{\mathcal{M}}^{\left( 4\right) }=\left\{ \lambda
_{0},\lambda _{1},\lambda _{2},\lambda _{3},\lambda _{4}\right\} $ as the
relevant abelian semigroup whose elements satisfy the multiplication law%
\begin{equation}
\lambda _{\alpha }\lambda _{\beta }=\left\{
\begin{array}{c}
\lambda _{\alpha +\beta },\text{ \ \ \ \ if }\alpha +\beta \leq 4\text{ } \\
\lambda _{\alpha +\beta -4},\text{ \ if }\alpha +\beta >4\text{\ }%
\end{array}%
\right.
\end{equation}%
Let $S_{\mathcal{M}}^{\left( 4\right) }=S_{0}\cup S_{1}\cup S_{2}$ be a
subset decomposition with%
\begin{eqnarray*}
S_{0} &=&\left\{ \lambda _{0},\lambda _{2},\lambda _{4}\right\} , \\
S_{1} &=&\left\{ \lambda _{1},\lambda _{3}\right\} \,, \\
S_{2} &=&\left\{ \lambda _{2},\lambda _{4}\right\} \,,
\end{eqnarray*}%
where $S_{0}$, $S_{1}$ and $S_{2}$ satisfy the resonance condition [compare
with Eqs. (\ref{SD1})-(\ref{SD2})]%
\begin{eqnarray}
S_{0}\cdot S_{0} &\subset &S_{0}\,,\text{ \ \ \ \ }S_{1}\cdot S_{1}\subset
S_{0}\cap S_{2}\,, \\
S_{0}\cdot S_{1} &\subset &S_{1}\,,\text{ \ \ \ \ }S_{1}\cdot S_{2}\subset
S_{1}\,, \\
S_{0}\cdot S_{2} &\subset &S_{2}\,,\text{ \ \ \ \ }S_{2}\cdot S_{2}\subset
S_{0}\,.
\end{eqnarray}%
Then according to Ref.~\cite{Sexp},%
\begin{equation*}
\mathfrak{G}_{R}=W_{0}\oplus W_{1}\oplus W_{2}\,,
\end{equation*}%
is a resonant subalgebra of $S_{\mathcal{M}}^{\left( 4\right) }\times
\mathfrak{osp}\left( 2|2\right) \otimes \mathfrak{sp}\left( 2\right) $ where%
\begin{eqnarray*}
W_{0} &=&\left( S_{0}\times V_{0}\right) =\left\{ \lambda _{0},\lambda
_{2},\lambda _{4}\right\} \times \left\{ \tilde{J}_{ab},\tilde{T}%
^{ij}\right\} \,, \\
W_{1} &=&\left( S_{1}\times V_{1}\right) =\left\{ \lambda _{1},\lambda
_{3}\right\} \times \left\{ \tilde{Q}_{\alpha }^{i}\right\} \,, \\
W_{2} &=&\left( S_{2}\times V_{2}\right) =\left\{ \lambda _{2},\lambda
_{4}\right\} \times \left\{ \tilde{P}_{a}\right\} \,.
\end{eqnarray*}%
The expanded superalgebra corresponds to a $\left( 2,0\right) $ $AdS$%
-Lorentz superalgebra and is generated by the set of generators $\left\{
J_{ab},P_{a},\tilde{Z}_{ab},\tilde{Z}_{a},Z_{ab},T^{ij},\tilde{Y}%
^{ij},Y^{ij},Q_{\alpha }^{i},\Sigma _{\alpha }^{i}\right\} $ which are
related to the original ones through%
\begin{eqnarray*}
J_{ab} &=&\lambda _{0}\tilde{J}_{ab}\,,\text{ \ \ \ }P_{a}=\lambda _{2}%
\tilde{P}_{a}\,,\text{ \ \ \ }Q_{\alpha }^{i}=\lambda _{1}\tilde{Q}_{\alpha
}^{i}\,, \\
\tilde{Z}_{ab} &=&\lambda _{2}\tilde{J}_{ab}\,,\text{ \ \ \ }\tilde{Z}%
_{a}=\lambda _{4}\tilde{P}_{a}\,,\text{ \ \ \ }\Sigma _{\alpha }^{i}=\lambda
_{3}\tilde{Q}_{\alpha }^{i}\,, \\
Z_{ab} &=&\lambda _{4}\tilde{J}_{ab}\,,\text{ \ \ \ }T^{ij}=\lambda _{0}%
\tilde{T}^{ij}\,,\text{ \ \ \ }\tilde{Y}^{ij}=\lambda _{2}\tilde{T}^{ij}\,,
\\
Y^{ij} &=&\lambda _{4}\tilde{T}^{ij}\,.
\end{eqnarray*}%
The explicit (anti)commutation relations can be derived using the
multiplication law of the semigroup and the original superalgebra (the $%
\mathcal{N}$-extended $AdS$-Lorentz superalgebra can be found in Appendix A).

In order to construct the explicit supergravity action let us first derive
the invariant tensor for the $\left( 2,0\right) $ $AdS$-Lorentz
superalgebra. According to Theorem VII.2 of Ref.~\cite{Sexp}, it is possible
to show that the non-vanishing components of the invariant tensor for the $%
\mathcal{N}=2$ $AdS$-Lorentz are, besides those given by eqs.(\ref{invt3a})-(%
\ref{invt3c}),%
\begin{eqnarray}
\left\langle Q_{\alpha }^{i}Q_{\beta }^{j}\right\rangle &=&\left\langle
\Sigma _{\alpha }^{i}\Sigma _{\beta }^{j}\right\rangle =\tilde{\alpha}%
_{2}\left\langle \tilde{Q}_{\alpha }^{i}\tilde{Q}_{\beta }^{j}\right\rangle
=2\left( \beta _{2}-\alpha _{2}\right) C_{\alpha \beta }\,\delta ^{ij}\,,
\label{Ninvta} \\
\left\langle Q_{\alpha }^{i}\Sigma _{\beta }^{j}\right\rangle &=&\tilde{%
\alpha}_{4}\left\langle \tilde{Q}_{\alpha }^{i}\tilde{Q}_{\beta
}^{j}\right\rangle =2\left( \beta _{4}-\alpha _{4}\right) C_{\alpha \beta
}\,\delta ^{ij}, \\
\left\langle T^{ij}T^{kl}\right\rangle &=&\tilde{\alpha}_{0}\left\langle
\tilde{T}^{ij}\tilde{T}^{kl}\right\rangle =2\left( \alpha _{0}-\beta
_{0}\right) \left( \delta ^{il}\delta ^{kj}-\delta ^{ik}\delta ^{lj}\right)
\,, \\
\left\langle T^{ij}\tilde{Y}^{kl}\right\rangle &=&\left\langle \tilde{Y}%
^{ij}Y^{kl}\right\rangle =\tilde{\alpha}_{2}\left\langle \tilde{T}^{ij}%
\tilde{T}^{kl}\right\rangle =2\left( \alpha _{2}-\beta _{2}\right) \left(
\delta ^{il}\delta ^{kj}-\delta ^{ik}\delta ^{lj}\right) \,, \\
\left\langle T^{ij}Y^{kl}\right\rangle &=&\left\langle \tilde{Y}^{ij}\tilde{Y%
}^{kl}\right\rangle =\left\langle Y^{ij}Y^{kl}\right\rangle =\tilde{\alpha}%
_{4}\left\langle \tilde{T}^{ij}\tilde{T}^{kl}\right\rangle =2\left( \alpha
_{4}-\beta _{4}\right) \left( \delta ^{il}\delta ^{kj}-\delta ^{ik}\delta
^{lj}\right) \,,  \label{Ninvtb}
\end{eqnarray}%
where $\left\{ \tilde{J}_{ab},\tilde{P}_{a},\tilde{T}^{ij},\tilde{Q}_{\alpha
}^{i}\right\} $ generate the $\mathfrak{osp}\left( 2|2\right) \otimes
\mathfrak{sp}\left( 2\right) $ superalgebra and where we have defined%
\begin{align*}
\alpha _{0}& \equiv \tilde{\alpha}_{0}\mu _{0}\,,\text{ \ \ \ \ }\alpha
_{2}\equiv \tilde{\alpha}_{2}\mu _{0}\,,\text{ \ \ \ \ }\alpha _{4}\equiv
\tilde{\alpha}_{4}\mu _{0}\,, \\
\beta _{0}& \equiv \tilde{\alpha}_{0}\mu _{1}\,,\text{ \ \ \ \ }\beta
_{2}\equiv \tilde{\alpha}_{2}\mu _{1}\,,\text{ \ \ \ \ }\beta _{4}\equiv
\tilde{\alpha}_{4}\mu _{1}\,.\text{\ }
\end{align*}%
Here $\tilde{\alpha}_{0},\tilde{\alpha}_{2},\tilde{\alpha}_{4}$ are
arbitrary constants as well as $\mu _{0}$ and $\mu _{1}.$To construct the CS
supergravity action we require, in addition to the invariant tensor, the
gauge connection one-form:%
\begin{eqnarray}
A &=&\frac{1}{2}\omega ^{ab}J_{ab}+\frac{1}{2}\tilde{k}^{ab}\tilde{Z}_{ab}+%
\frac{1}{2}k^{ab}Z_{ab}+\frac{1}{l}e^{a}P_{a}+\frac{1}{l}\tilde{h}^{a}\tilde{%
Z}_{a}  \notag \\
&&+\frac{1}{2}A^{ij}T_{ij}+\frac{1}{2}\tilde{B}^{ij}\tilde{Y}_{ij}+\frac{1}{2%
}B^{ij}Y_{ij}+\frac{1}{\sqrt{l}}\bar{\psi}_{i}Q^{i}+\frac{1}{\sqrt{l}}\bar{%
\xi}_{i}\Sigma ^{i}\,.  \label{N2OF}
\end{eqnarray}%
The associate curvature two-form $F=dA+AA$ is given by%
\begin{eqnarray}
F &=&\frac{1}{2}R^{ab}J_{ab}+\frac{1}{2}\tilde{F}^{ab}\tilde{Z}_{ab}+\frac{1%
}{2}F^{ab}Z_{ab}+\frac{1}{l}F^{a}P_{a}+\frac{1}{l}\tilde{H}^{a}\tilde{Z}_{a}
\notag \\
&&+\frac{1}{2}F^{ij}T_{ij}+\frac{1}{2}\tilde{G}^{ij}\tilde{Y}_{ij}+\frac{1}{2%
}G^{ij}Y_{ij}+\frac{1}{\sqrt{l}}\bar{\Psi}_{i}Q^{i}+\frac{1}{\sqrt{l}}\bar{%
\Xi}_{i}\Sigma ^{i}\,,  \label{N2TF}
\end{eqnarray}%
where%
\begin{align*}
R^{ab}& =d\omega ^{ab}+\omega _{\text{ }c}^{a}\omega ^{cb}\,, \\
F^{a}& =de^{a}+\omega _{\text{ }b}^{a}e^{b}+k_{\text{ }b}^{a}e^{b}+\tilde{k}%
_{\text{ }b}^{a}\tilde{h}^{b}-\frac{1}{2}\bar{\psi}^{i}\Gamma ^{a}\psi ^{i}-%
\frac{1}{2}\bar{\xi}^{i}\Gamma ^{a}\xi ^{i}\,, \\
\tilde{H}^{a}& =d\tilde{h}^{a}+\omega _{\text{ }b}^{a}\tilde{h}^{b}+\tilde{k}%
_{\text{ }b}^{a}e^{b}+k_{\text{ }b}^{a}\tilde{h}^{b}-\bar{\psi}^{i}\Gamma
^{a}\xi ^{i}\,, \\
\tilde{F}^{ab}& =d\tilde{k}^{ab}+\omega _{\text{ }c}^{a}\tilde{k}%
^{cb}-\omega _{\text{ }c}^{b}\tilde{k}^{ca}+k_{\text{ }c}^{a}\tilde{k}%
^{cb}-k_{\text{ }c}^{b}\tilde{k}^{ca}+\frac{2}{l^{2}}e^{a}\tilde{h}^{b}+%
\frac{1}{2l}\bar{\psi}^{i}\Gamma ^{ab}\psi ^{i}+\frac{1}{2l}\bar{\xi}%
^{i}\Gamma ^{ab}\xi ^{i}\,, \\
F^{ab}& =dk^{ab}+\omega _{\text{ }c}^{a}k^{cb}-\omega _{\text{ }c}^{b}k^{ca}+%
\tilde{k}_{\text{ }c}^{a}\tilde{k}^{cb}+k_{\text{ }c}^{a}k^{cb}+\frac{1}{%
l^{2}}e^{a}e^{b}+\frac{1}{l^{2}}\tilde{h}^{a}\tilde{h}^{b}+\frac{1}{l}\bar{%
\xi}^{i}\Gamma ^{ab}\psi ^{i}\,,
\end{align*}%
\begin{align*}
F^{ij}& =dA^{ij}+A^{ik}A^{kj}\,, \\
\tilde{G}^{ij}& =d\tilde{B}^{ij}+A^{ik}\tilde{B}^{kj}+\tilde{B}^{ik}A^{kj}+%
\tilde{B}^{ik}B^{kj}+B^{ik}\tilde{B}^{kj}+\bar{\psi}^{i}\psi ^{j}\,+\bar{\xi}%
^{i}\xi ^{j}, \\
G^{ij}& =dB^{ij}+A^{ik}B^{kj}+B^{ik}A^{kj}+\tilde{B}^{ik}\tilde{B}%
^{kj}+B^{ik}B^{kj}+2\bar{\psi}^{i}\xi ^{j}\,, \\
\Psi ^{i}& =d\psi ^{i}+\frac{1}{4}\omega _{ab}\Gamma ^{ab}\psi ^{i}+\frac{1}{%
4}k_{ab}\Gamma ^{ab}\psi ^{i}+\frac{1}{4}\tilde{k}_{ab}\Gamma ^{ab}\xi ^{i}+%
\frac{1}{2l}e_{a}\Gamma ^{a}\xi ^{i}+\frac{1}{2l}\tilde{h}_{a}\Gamma
^{a}\psi ^{i} \\
& +A^{ij}\psi ^{j}+B^{ij}\psi ^{j}+\tilde{B}^{ij}\xi ^{j}, \\
\Xi ^{i}& =d\xi ^{i}+\frac{1}{4}\omega _{ab}\Gamma ^{ab}\xi ^{i}+\frac{1}{4}%
k_{ab}\Gamma ^{ab}\xi ^{i}+\frac{1}{4}\tilde{k}_{ab}\Gamma ^{ab}\psi ^{i}+%
\frac{1}{2l}e_{a}\Gamma ^{a}\psi ^{i}+\frac{1}{2l}\tilde{h}_{a}\Gamma
^{a}\xi ^{i}\, \\
& +A^{ij}\xi ^{j}+B^{ij}\xi ^{j}+\tilde{B}^{ij}\psi ^{j}\,.
\end{align*}%
Then considering the connection one-form (\ref{N2OF}) and the non-vanishing
components of the invariant tensor ((\ref{invt3a})-(\ref{invt3c}) and (\ref%
{Ninvta})-(\ref{Ninvtb})) in the general three-dimensional CS expression, we
find the $\left( 2,0\right) $ $AdS$-Lorentz CS action supergravity up to a
surface term:

\bigskip
\begin{eqnarray}
I_{CS}^{\left( 2+1\right) } &=&k\int \frac{\alpha _{0}}{2}\left( \omega _{%
\text{ }b}^{a}d\omega _{\text{ }a}^{b}+\frac{2}{3}\omega _{\text{ }%
c}^{a}\omega _{\text{ }b}^{c}\omega _{\text{ }a}^{b}\right) +\left( \alpha
_{0}-\beta _{0}\right) \left( A^{ij}dA^{ji}+\frac{2}{3}A^{ik}A^{kj}A^{ji}%
\right)  \notag \\
&&+\frac{\beta _{2}}{l}\epsilon _{abc}\left( R^{ab}e^{c}+\frac{1}{3l^{2}}%
e^{a}e^{b}e^{c}+K^{ab}e^{c}+\tilde{K}^{ab}\tilde{h}^{c}+\frac{1}{l^{2}}%
\tilde{h}^{a}\tilde{h}^{b}e\right)  \notag \\
&&+\left( \alpha _{2}-\beta _{2}\right) \left[ \tilde{B}^{ij}\left(
dA^{ji}+A^{jk}A^{ki}\right) +\tilde{B}^{ij}\left(
dB^{ji}+A^{jk}B^{ki}+B^{jk}A^{ki}+\tilde{B}^{jk}\tilde{B}^{ki}+B^{jk}B^{ki}%
\right) \right.  \notag \\
&&\left. +\left( A^{ij}+B^{ij}\right) \left( d\tilde{B}^{ji}+A^{jk}\tilde{B}%
^{ki}+\tilde{B}^{jk}A^{ki}+\tilde{B}^{jk}B^{ki}+B^{jk}\tilde{B}^{ki}\right) -%
\frac{2}{l}\bar{\psi}^{i}\Psi ^{i}-\frac{2}{l}\bar{\xi}^{i}\Xi ^{i}\right]
\notag \\
&&+\alpha _{2}\left[ R_{\text{ }b}^{a}\tilde{k}_{\text{ }a}^{b}+K_{\text{ }%
b}^{a}\tilde{k}_{\text{ }a}^{b}+\tilde{K}_{\text{ }b}^{a}k_{\text{ }a}^{b}+%
\frac{1}{l^{2}}e^{a}H_{a}+\frac{1}{l^{2}}\tilde{h}^{a}K_{a}\right]  \notag \\
&&+\frac{\beta _{4}}{l}\epsilon _{abc}\left( R^{ab}\tilde{h}^{c}+\frac{1}{%
3l^{2}}\tilde{h}^{a}\tilde{h}^{b}\tilde{h}^{c}+\tilde{K}^{ab}e^{c}+K^{ab}%
\tilde{h}^{c}+\frac{1}{l^{2}}e^{a}e^{b}\tilde{h}^{c}\right)  \notag \\
&&+\left( \alpha _{4}-\beta _{4}\right) \left[ \tilde{B}^{ij}\left( d\tilde{B%
}^{ji}+A^{jk}\tilde{B}^{ki}+\tilde{B}^{jk}A^{ki}+\tilde{B}^{jk}B^{ki}+B^{jk}%
\tilde{B}^{ki}\right) +B^{ij}\left( dA^{ji}+A^{jk}A^{ki}\right) \right.
\notag \\
&&\left. +\left( A^{ij}+B^{ij}\right) \left(
dB^{ji}+A^{jk}B^{ki}+B^{jk}A^{ki}+\tilde{B}^{jk}\tilde{B}^{ki}+B^{jk}B^{ki}%
\right) -\frac{2}{l}\bar{\xi}^{i}\Psi ^{i}-\frac{2}{l}\bar{\psi}^{i}\Xi ^{i}%
\right]  \notag \\
&&+\alpha _{4}\left[ R_{\text{ }b}^{a}k_{\text{ }a}^{b}+K_{\text{ }b}^{a}k_{%
\text{ }a}^{b}+\tilde{K}_{\text{ }b}^{a}\tilde{k}_{\text{ }a}^{b}+\frac{1}{%
l^{2}}e^{a}K_{a}+\frac{1}{l^{2}}\tilde{h}^{a}H_{a}\right] \,,  \label{N2AL}
\end{eqnarray}%
where $\Psi ^{i}$ and $\Xi ^{i}$ are the fermionic components of the
curvature two form and%
\begin{eqnarray*}
K^{ab} &=&Dk^{ab}+k_{\text{ }d}^{a}k_{\text{ }b}^{d}+\tilde{k}_{\text{ }%
d}^{a}\tilde{k}_{\text{ }b}^{d}\,,\text{ \ \ \ }\tilde{K}^{ab}=D\tilde{k}%
^{ab}+k_{\text{ }d}^{a}\tilde{k}_{\text{ }b}^{d}+k_{b}^{\text{ }d}\tilde{k}_{%
\text{ }d}^{a}\,, \\
H^{a} &=&D\tilde{h}^{a}+k_{\text{ }b}^{a}\tilde{h}^{b}+\tilde{k}_{\text{ }%
b}^{a}e^{b}\,,\text{ \ \ \ \ \ \ }K^{a}=T^{a}+k_{\text{ }b}^{a}e^{b}+\tilde{k%
}_{\text{ }b}^{a}\tilde{h}^{b}\,,
\end{eqnarray*}%
As in the $\mathcal{N}=1$ case, the $\left( 2,0\right) $ Maxwell
supergravity cannot be trivially obtained considering the rescaling of the
fields in (\ref{N2AL}) and applying some limit.

\subsection{$\mathcal{N}=2$ Maxwell Supergravity}

As the $\mathfrak{osp}\left( 2|2\right) \otimes \mathfrak{sp}\left( 2\right)
$ superalgebra has its $\left( 2,0\right) $ Poincar\'{e} limit, the $\left(
2,0\right) $ $AdS$-Lorentz superalgebra possesses its proper IW contracted
superalgebra. Indeed, after rescaling the generators%
\begin{align*}
\tilde{Z}_{ab}& \rightarrow \sigma ^{2}\tilde{Z}_{ab}\,,\text{ \ \ }%
Z_{ab}\rightarrow \sigma ^{4}Z_{ab}\,,\text{ \ \ }P_{a}\rightarrow \sigma
^{2}P_{a}\,,\text{ \ \ \ \ }J_{ab}\rightarrow J_{ab}\,, \\
\tilde{Z}_{a}& \rightarrow \sigma ^{4}\tilde{Z}_{a}\,,\text{ \ \ }Q_{\alpha
}^{i}\rightarrow \sigma Q_{\alpha }^{i}\,,\text{\ \ }\Sigma ^{i}\rightarrow
\sigma ^{3}\Sigma ^{i}\,,\text{ \ \ }Y^{ij}\rightarrow \sigma ^{4}Y^{ij}\,,
\\
\tilde{Y}^{ij}& \rightarrow \sigma ^{2}\tilde{Y}^{ij}\,,\text{ \ \ }%
T^{ij}\rightarrow T^{ij}\,,
\end{align*}%
and applying the limit $\sigma \rightarrow \infty $, we obtain the $\mathcal{%
N}=2$ Maxwell superalgebra whose (anti)commutation relations can be found in
Refs.~\cite{AILW, CR1} (see Appendix B for $p=2,$ $q=0$).

As in the previous case, the CS supergravity action for the $\left(
2,0\right) $ Maxwell supergroup can be derived combining the IW contraction
with the $S$-expanded invariant tensor. Indeed, it is necessary to extend
the rescaling of the generators to the\ $\alpha $ and $\beta $ constants
appearing in the non-vanishing components of the invariant tensor of the $%
\left( 2,0\right) $ $AdS$-Lorentz superalgebra. A rescaling which preserves
the curvatures structure is given by%
\begin{eqnarray*}
\alpha _{4} &\rightarrow &\sigma ^{4}\alpha _{4}\,,\text{ \ \ \ \ }\alpha
_{2}\rightarrow \sigma ^{2}\alpha _{2}\,,\text{ \ \ \ \ }\alpha
_{0}\rightarrow \alpha _{0}\,, \\
\beta _{4} &\rightarrow &\sigma ^{4}\beta _{4}\,,\text{ \ \ \ \ }\,\beta
_{2}\rightarrow \sigma ^{2}\beta _{2}\,,\text{ \ \ \ \ }\beta
_{0}\rightarrow \beta _{0}\,.
\end{eqnarray*}%
Then, considering the rescaling of both constants and generators, and
applying the limit $\sigma \rightarrow \infty $, we obtain the $\left(
2,0\right) $ Maxwell non-vanishing components of the invariant tensor,%
\begin{align}
\left\langle J_{ab}J_{cd}\right\rangle _{\mathcal{M}}& =\alpha _{0}\left(
\eta _{ad}\eta _{bc}-\eta _{ac}\eta _{bd}\right) \,, \\
\text{\ \ }\left\langle J_{ab}\tilde{Z}_{cd}\right\rangle _{\mathcal{M}}&
=\alpha _{2}\left( \eta _{ad}\eta _{bc}-\eta _{ac}\eta _{bd}\right) \,, \\
\left\langle \tilde{Z}_{ab}\tilde{Z}_{cd}\right\rangle _{\mathcal{M}}&
=\left\langle J_{ab}Z_{cd}\right\rangle =\alpha _{4}\left( \eta _{ad}\eta
_{bc}-\eta _{ac}\eta _{bd}\right) \,, \\
\left\langle J_{ab}P_{c}\right\rangle _{\mathcal{M}}& =\beta _{2}\epsilon
_{abc}\,, \\
\left\langle \tilde{Z}_{ab}P_{c}\right\rangle _{\mathcal{M}}& =\left\langle
J_{ab}\tilde{Z}_{c}\right\rangle _{\mathcal{M}}=\beta _{4}\epsilon _{abc}\,,
\\
\left\langle P_{a}P_{b}\right\rangle _{\mathcal{M}}& =\alpha _{4}\eta
_{ab}\,,
\end{align}%
\begin{eqnarray}
\left\langle Q_{\alpha }^{i}Q_{\beta }^{j}\right\rangle _{\mathcal{M}}
&=&2\left( \beta _{2}-\alpha _{2}\right) C_{\alpha \beta }\,\delta ^{ij}\,,
\\
\left\langle Q_{\alpha }^{i}\Sigma _{\beta }^{j}\right\rangle _{\mathcal{M}}
&=&2\left( \beta _{4}-\alpha _{4}\right) C_{\alpha \beta }\,\delta ^{ij}, \\
\left\langle T^{ij}T^{kl}\right\rangle _{\mathcal{M}} &=&2\left( \alpha
_{0}-\beta _{0}\right) \left( \delta ^{il}\delta ^{kj}-\delta ^{ik}\delta
^{lj}\right) \,, \\
\left\langle T^{ij}\tilde{Y}^{kl}\right\rangle _{\mathcal{M}} &=&2\left(
\alpha _{2}-\beta _{2}\right) \left( \delta ^{il}\delta ^{kj}-\delta
^{ik}\delta ^{lj}\right) \,, \\
\left\langle T^{ij}Y^{kl}\right\rangle _{\mathcal{M}} &=&\left\langle \tilde{%
Y}^{ij}\tilde{Y}^{kl}\right\rangle _{\mathcal{M}}=2\left( \alpha _{4}-\beta
_{4}\right) \left( \delta ^{il}\delta ^{kj}-\delta ^{ik}\delta ^{lj}\right)
\,,
\end{eqnarray}%
where $\alpha _{0},\alpha _{2},\alpha _{4},\beta _{2}$ and $\beta _{4}$ are
arbitrary constants and the generators now satisfy the $\left( 2,0\right) $
Maxwell (anti)commutation relations. In order to write down a CS action we
require the gauge connection one-form given by%
\begin{eqnarray}
A &=&\frac{1}{2}\omega ^{ab}J_{ab}+\frac{1}{2}\tilde{k}^{ab}\tilde{Z}_{ab}+%
\frac{1}{2}k^{ab}Z_{ab}+\frac{1}{l}e^{a}P_{a}+\frac{1}{l}\tilde{h}^{a}\tilde{%
Z}_{a}  \notag \\
&&+\frac{1}{2}A^{ij}T_{ij}+\frac{1}{2}\tilde{B}^{ij}\tilde{Y}_{ij}+\frac{1}{2%
}B^{ij}Y_{ij}+\frac{1}{\sqrt{l}}\bar{\psi}_{i}Q^{i}+\frac{1}{\sqrt{l}}\bar{%
\xi}_{i}\Sigma ^{i}\,.
\end{eqnarray}%
Considering the non-vanishing components of the invariant tensor, the CS
action for the $\mathcal{N}=2$ Maxwell superalgebra reduces to

\bigskip
\begin{eqnarray}
I_{\mathcal{M}-CS}^{\left( 2+1\right) } &=&k\int \frac{\alpha _{0}}{2}\left(
\omega _{\text{ }b}^{a}d\omega _{\text{ }a}^{b}+\frac{2}{3}\omega _{\text{ }%
c}^{a}\omega _{\text{ }b}^{c}\omega _{\text{ }a}^{b}\right) +\left( \alpha
_{0}-\beta _{0}\right) \left( A^{ij}dA^{ji}+\frac{2}{3}A^{ik}A^{kj}A^{ji}%
\right)   \notag \\
&&+\frac{\beta _{2}}{l}\epsilon _{abc}R^{ab}e^{c}+\alpha _{2}R_{\text{ }%
b}^{a}\tilde{k}_{\text{ }a}^{b}  \notag \\
&&+\left( \alpha _{2}-\beta _{2}\right) \left[ \tilde{B}^{ij}\left(
dA^{ji}+A^{jk}A^{ki}\right) +A^{ij}\left( d\tilde{B}^{ji}+A^{jk}\tilde{B}%
^{ki}+\tilde{B}^{jk}A^{ki}\right) -\frac{2}{l}\bar{\psi}^{i}\Psi ^{i}\right]
\notag \\
&&+\frac{\beta _{4}}{l}\epsilon _{abc}\left( R^{ab}\tilde{h}^{c}+D_{\omega }%
\tilde{k}^{ab}e^{c}\right)   \notag \\
&&+\left( \alpha _{4}-\beta _{4}\right) \left[ \tilde{B}^{ij}\left( d\tilde{B%
}^{ji}+A^{jk}\tilde{B}^{ki}+\tilde{B}^{jk}A^{ki}\right) +B^{ij}\left(
dA^{ji}+A^{jk}A^{ki}\right) \right.   \notag \\
&&\left. +\left( A^{ij}\right) \left( dB^{ji}+A^{jk}B^{ki}+B^{jk}A^{ki}+%
\tilde{B}^{jk}\tilde{B}^{ki}\right) -\frac{2}{l}\bar{\xi}^{i}\Psi ^{i}-\frac{%
2}{l}\bar{\psi}^{i}\Xi ^{i}\right]   \notag \\
&&+\alpha _{4}\left[ R_{\text{ }b}^{a}k_{\text{ }a}^{b}+D_{\omega }\tilde{k}%
_{\text{ }b}^{a}\tilde{k}_{\text{ }a}^{b}+\frac{1}{l^{2}}e^{a}T_{a}\right]
\,,
\end{eqnarray}%
where%
\begin{eqnarray*}
\Psi ^{i} &=&d\psi ^{i}+\frac{1}{4}\omega _{ab}\Gamma ^{ab}\psi
^{i}+A^{ij}\psi ^{j}\,, \\
\Xi ^{i} &=&d\xi ^{i}+\frac{1}{4}\omega _{ab}\Gamma ^{ab}\xi ^{i}+\frac{1}{4}%
\tilde{k}_{ab}\Gamma ^{ab}\psi ^{i}+\frac{1}{2l}e_{a}\Gamma ^{a}\psi
^{i}+A^{ij}\xi ^{j}+\tilde{B}^{ij}\psi ^{j}\,.
\end{eqnarray*}%
This CS supergravity action is invariant up to boundary terms under the
local gauge transformations of the $\mathcal{N}=2$ Maxwell supergroup. In
particular, under the supersymmetric transformations, the fields transform as%
\begin{align}
\delta \omega ^{ab}& =0\,,\text{ \ \ \ \ \ }\delta e^{a}=\bar{\epsilon}%
^{i}\Gamma ^{a}\psi ^{i},  \label{ST01} \\
\delta \tilde{k}^{ab}& =-\frac{1}{l}\bar{\epsilon}^{i}\Gamma ^{ab}\psi ^{i},
\\
\delta k^{ab}& =-\frac{1}{l}\bar{\varrho}^{i}\Gamma ^{ab}\psi ^{i}-\frac{1}{l%
}\bar{\epsilon}^{i}\Gamma ^{ab}\xi ^{i},
\end{align}%
\begin{align}
\delta \tilde{h}^{a}& =\,\bar{\varrho}^{i}\Gamma ^{a}\psi ^{i}+\bar{\epsilon}%
^{i}\Gamma ^{a}\xi ^{i}\text{\thinspace }, \\
\delta A^{ij}& =0\,, \\
\delta \tilde{B}^{ij}& =-\frac{2}{l}\bar{\psi}^{\left[ i\right. }\epsilon
^{\left. j\right] }\,, \\
\delta B^{ij}& =-\frac{2}{l}\bar{\psi}^{\left[ i\right. }\varrho ^{\left. j%
\right] }\,, \\
\delta \psi ^{i}& =d\epsilon ^{i}+\frac{1}{4}\omega ^{ab}\Gamma
_{ab}\epsilon ^{i}+A^{ij}\epsilon ^{j}, \\
\delta \xi ^{i}& =d\varrho ^{i}+\frac{1}{4}\omega ^{ab}\Gamma _{ab}\varrho
^{i}+\frac{1}{2l}e^{a}\Gamma _{a}\epsilon ^{i}+\frac{1}{4}\tilde{k}%
^{ab}\Gamma _{ab}\epsilon ^{i}\, \\
& +A^{ij}\varrho ^{j}+\tilde{B}^{ij}\epsilon ^{j}\,,
\end{align}%
where the $\epsilon ^{i}$ and $\varrho ^{i}$ parameters are related to the $%
Q^{i}$ and $\Sigma ^{i}$ generators, respectively.

We remark that the generalized cosmological constant term appearing in the $%
\left( 2,0\right) $ $AdS$-Lorentz supergravity model is no longer present
after the IW contraction. This is analogous to the Poincar\'{e} limit from
the $AdS$ one. However, unlike the Poincar\'{e} supergravity theory, the
internal symmetry fields appear explicitly in the exotic Lagrangian.
Additionally, the spinorial fields contribute to the exotic like part.

\subsection{$\left( p,q\right) $ $AdS$-Lorentz Supergravity and the Maxwell
limit}

In this section we present the three-dimensional $\mathcal{N}=p+q$ extended $%
AdS$-Lorentz Supergravity and its Maxwell limit applying the IW contraction
not only at the generators level but also to the constants appearing in the
invariant tensor. To this purpose we expand the three-dimensional $\left(
p,q\right) $ exotic supergravity theory \cite{GTW}, in order to obtain the
local $AdS$-Lorentz supersymmetric extension. In particular, we generalize
the Poincar\'{e} limit showed in section 2 to the Maxwell limit.

The $\left( p,q\right) $ $AdS$-Lorentz superalgebra can be obtained as an $S$%
-expansion of the $\mathfrak{osp}\left( 2|p\right) \otimes \mathfrak{osp}%
\left( 2|q\right) $ superalgebra. Indeed, considering $S_{\mathcal{M}%
}^{\left( 4\right) }=\left\{ \lambda _{0},\lambda _{1},\lambda _{2},\lambda
_{3},\lambda _{4}\right\} $ as the relevant semigroup whose elements satisfy%
\begin{equation*}
\lambda _{\alpha }\lambda _{\beta }=\left\{
\begin{array}{c}
\lambda _{\alpha +\beta },\text{ \ \ \ \ if }\alpha +\beta \leq 4\text{ } \\
\lambda _{\alpha +\beta -4},\text{ \ if }\alpha +\beta >4\text{\ }%
\end{array}%
\right.
\end{equation*}%
and considering the resonant condition (see $\mathcal{N}=2$ case), we obtain
a new superlagebra generated by $\left\{ J_{ab},P_{a},\tilde{Z}_{ab},\tilde{Z%
}_{a},Z_{ab},T^{ij},\tilde{Y}^{ij},Y^{ij},T^{IJ},\tilde{Y}%
^{IJ},Y^{IJ},Q_{\alpha }^{i},\Sigma _{\alpha }^{i},Q_{\alpha }^{I},\Sigma
_{\alpha }^{J}\right\} $ whose generators satisfy the $\left( p,q\right) $ $%
AdS$-Lorentz superalgebra. In particular, besides satisfying the
(anti)commutation relations appearing in Appendix A, the $I$-index
generators satisfy%
\begin{eqnarray}
\left[ T^{IJ},T^{KL}\right] &=&\delta ^{JK}T^{IL}-\delta ^{IK}T^{JL}-\delta
^{JL}T^{IK}+\delta ^{IL}T^{JK}\text{\thinspace }, \\
\left[ T^{IJ},\tilde{Y}^{KL}\right] &=&\delta ^{JK}\tilde{Y}^{IL}-\delta
^{IK}\tilde{Y}^{JL}-\delta ^{JL}\tilde{Y}^{IK}+\delta ^{IL}\tilde{Y}^{JK}%
\text{\thinspace }, \\
\left[ T^{IJ},Y^{KL}\right] &=&\delta ^{JK}Y^{IL}-\delta ^{IK}Y^{JL}-\delta
^{JL}Y^{IK}+\delta ^{IL}Y^{JK}\text{\thinspace }, \\
\left[ \tilde{Y}^{IJ},\tilde{Y}^{KL}\right] &=&\delta ^{JK}Y^{IL}-\delta
^{IK}Y^{JL}-\delta ^{JL}Y^{IK}+\delta ^{IL}Y^{JK}\text{\thinspace }, \\
\left[ \tilde{Y}^{IJ},Y^{KL}\right] &=&\delta ^{JK}\tilde{Y}^{IL}-\delta
^{IK}\tilde{Y}^{JL}-\delta ^{JL}\tilde{Y}^{IK}+\delta ^{IL}\tilde{Y}^{JK}\,,
\\
\left[ Y^{IJ},Y^{KL}\right] &=&\delta ^{JK}Y^{IL}-\delta ^{IK}Y^{JL}-\delta
^{JL}Y^{IK}+\delta ^{IL}Y^{JK}\text{\thinspace },
\end{eqnarray}%
\begin{align}
\left[ J_{ab},Q_{\alpha }^{I}\right] & =-\frac{1}{2}\left( \Gamma
_{ab}Q^{I}\right) _{\alpha }\,,\text{ \ \ \ \ }\left[ P_{a},Q_{\alpha }^{I}%
\right] =\frac{1}{2}\left( \Gamma _{a}\Sigma ^{I}\right) _{\alpha }\,, \\
\left[ \tilde{Z}_{ab},Q_{\alpha }^{I}\right] & =-\frac{1}{2}\left( \Gamma
_{ab}\Sigma ^{I}\right) _{\alpha }\,,\text{ \ \ \ \ }\left[ \tilde{Z}%
_{a},Q_{\alpha }^{I}\right] =\frac{1}{2}\left( \Gamma _{a}Q^{I}\right)
_{\alpha }\,, \\
\left[ Z_{ab},Q_{\alpha }^{I}\right] & =-\frac{1}{2}\left( \Gamma
_{ab}Q^{I}\right) _{\alpha }\,,\text{ \ \ \ \ }\left[ P_{a},\Sigma _{\alpha
}^{I}\right] =\frac{1}{2}\left( \Gamma _{a}Q^{I}\right) _{\alpha }\,, \\
\left[ J_{ab},\Sigma _{\alpha }^{I}\right] & =-\frac{1}{2}\left( \Gamma
_{ab}\Sigma ^{I}\right) _{\alpha }\,,\text{ \ \ \ \ }\left[ \tilde{Z}%
_{a},\Sigma _{\alpha }^{I}\right] =\frac{1}{2}\left( \Gamma _{a}\Sigma
^{I}\right) _{\alpha }\,, \\
\left[ \tilde{Z}_{ab},\Sigma _{\alpha }^{I}\right] & =-\frac{1}{2}\left(
\Gamma _{ab}Q^{I}\right) _{\alpha }\,,\text{ \ \ \ \ }\left[ Z_{ab},\Sigma
_{\alpha }^{I}\right] =-\frac{1}{2}\left( \Gamma _{ab}\Sigma ^{I}\right)
_{\alpha }\,, \\
\left[ T^{IJ},Q_{\alpha }^{K}\right] & =(\delta ^{JK}Q_{\alpha }^{I}-\delta
^{IK}Q_{\alpha }^{J})\,,\text{ \ \ }\left[ \tilde{Y}^{IJ},Q_{\alpha }^{K}%
\right] =(\delta ^{JK}\Sigma _{\alpha }^{I}-\delta ^{IK}\Sigma _{\alpha
}^{J})\,,
\end{align}%
\begin{align}
\left[ Y^{IJ},Q_{\alpha }^{K}\right] & =(\delta ^{JK}Q_{\alpha }^{I}-\delta
^{IK}Q_{\alpha }^{J})\,,\text{ \ \ }\left[ T^{IJ},\Sigma _{\alpha }^{K}%
\right] =(\delta ^{JK}\Sigma _{\alpha }^{I}-\delta ^{IK}\Sigma _{\alpha
}^{J})\,, \\
\left[ \tilde{Y}^{IJ},\Sigma _{\alpha }^{K}\right] & =(\delta ^{JK}Q_{\alpha
}^{I}-\delta ^{IK}Q_{\alpha }^{J})\,,\text{ \ \ }\left[ Y^{IJ},\Sigma
_{\alpha }^{K}\right] =(\delta ^{JK}\Sigma _{\alpha }^{I}-\delta ^{IK}\Sigma
_{\alpha }^{J})\,, \\
\left\{ Q_{\alpha }^{I},Q_{\beta }^{J}\right\} & =\frac{1}{2}\delta ^{IJ}%
\left[ \left( \Gamma ^{ab}C\right) _{\alpha \beta }\tilde{Z}_{ab}+2\left(
\Gamma ^{a}C\right) _{\alpha \beta }P_{a}\right] \,-C_{\alpha \beta }\tilde{Y%
}^{IJ}, \\
\left\{ Q_{\alpha }^{I},\Sigma _{\beta }^{J}\right\} & =\frac{1}{2}\delta
^{IJ}\left[ \left( \Gamma ^{ab}C\right) _{\alpha \beta }Z_{ab}+2\left(
\Gamma ^{a}C\right) _{\alpha \beta }\tilde{Z}_{a}\right] -C_{\alpha \beta
}Y^{IJ}\,, \\
\left\{ \Sigma _{\alpha }^{I},\Sigma _{\beta }^{J}\right\} & =\frac{1}{2}%
\delta ^{IJ}\left[ \left( \Gamma ^{ab}C\right) _{\alpha \beta }\tilde{Z}%
_{ab}+2\left( \Gamma ^{a}C\right) _{\alpha \beta }P_{a}\right] \,-C_{\alpha
\beta }\tilde{Y}^{IJ}.
\end{align}%
Here, the $T^{ij},\tilde{Y}^{ij},Y^{ij},T^{IJ},\tilde{Y}^{IJ}$ and $Y^{IJ}$
generators correspond to internal symmetry generators with $i=1,\dots ,p~$%
and $I=1,\dots ,q$.

Using Theorem VII.2 of Ref.~\cite{Sexp}, it is possible to show that the
non-vanishing components of the invariant tensor for the $\mathcal{N}=p+q$ $%
AdS$-Lorentz are, besides those given by Eqs.(\ref{invt3a})-(\ref{invt3c}),%
\begin{eqnarray}
\left\langle Q_{\alpha }^{i}Q_{\beta }^{j}\right\rangle &=&\left\langle
\Sigma _{\alpha }^{i}\Sigma _{\beta }^{j}\right\rangle =\tilde{\alpha}%
_{2}\left\langle \tilde{Q}_{\alpha }^{i}\tilde{Q}_{\beta }^{j}\right\rangle
=2\left( \beta _{2}-\alpha _{2}\right) C_{\alpha \beta }\,\delta ^{ij}\,,
\label{pqinvt} \\
\left\langle Q_{\alpha }^{I}Q_{\beta }^{J}\right\rangle &=&\left\langle
\Sigma _{\alpha }^{I}\Sigma _{\beta }^{J}\right\rangle =\tilde{\alpha}%
_{2}\left\langle \tilde{Q}_{\alpha }^{I}\tilde{Q}_{\beta }^{J}\right\rangle
=2\left( \beta _{2}+\alpha _{2}\right) C_{\alpha \beta }\,\delta ^{IJ}\,, \\
\left\langle Q_{\alpha }^{i}\Sigma _{\beta }^{j}\right\rangle &=&\tilde{%
\alpha}_{4}\left\langle \tilde{Q}_{\alpha }^{i}\tilde{Q}_{\beta
}^{j}\right\rangle =2\left( \beta _{4}-\alpha _{4}\right) C_{\alpha \beta
}\,\delta ^{ij}, \\
\left\langle Q_{\alpha }^{I}\Sigma _{\beta }^{J}\right\rangle &=&\tilde{%
\alpha}_{4}\left\langle \tilde{Q}_{\alpha }^{I}\tilde{Q}_{\beta
}^{J}\right\rangle =2\left( \beta _{4}+\alpha _{4}\right) C_{\alpha \beta
}\,\delta ^{IJ},
\end{eqnarray}%
\begin{eqnarray}
\left\langle T^{ij}T^{kl}\right\rangle &=&\tilde{\alpha}_{0}\left\langle
\tilde{T}^{ij}\tilde{T}^{kl}\right\rangle =2\left( \alpha _{0}-\beta
_{0}\right) \left( \delta ^{il}\delta ^{kj}-\delta ^{ik}\delta ^{lj}\right)
\,, \\
\left\langle T^{IJ}T^{KL}\right\rangle &=&\tilde{\alpha}_{0}\left\langle
\tilde{T}^{IJ}\tilde{T}^{KL}\right\rangle =2\left( \alpha _{0}+\beta
_{0}\right) \left( \delta ^{IL}\delta ^{KJ}-\delta ^{IK}\delta ^{LJ}\right)
\,, \\
\left\langle T^{ij}\tilde{Y}^{kl}\right\rangle &=&\left\langle \tilde{Y}%
^{ij}Y^{kl}\right\rangle =\tilde{\alpha}_{2}\left\langle \tilde{T}^{ij}%
\tilde{T}^{kl}\right\rangle =2\left( \alpha _{2}-\beta _{2}\right) \left(
\delta ^{il}\delta ^{kj}-\delta ^{ik}\delta ^{lj}\right) \,, \\
\left\langle T^{IJ}\tilde{Y}^{KL}\right\rangle &=&\left\langle \tilde{Y}%
^{IJ}Y^{KL}\right\rangle =\tilde{\alpha}_{2}\left\langle \tilde{T}^{IJ}%
\tilde{T}^{KL}\right\rangle =2\left( \alpha _{2}+\beta _{2}\right) \left(
\delta ^{IL}\delta ^{KJ}-\delta ^{IK}\delta ^{LJ}\right) \,, \\
\left\langle T^{ij}Y^{kl}\right\rangle &=&\left\langle \tilde{Y}^{ij}\tilde{Y%
}^{kl}\right\rangle =\left\langle Y^{ij}Y^{kl}\right\rangle =\tilde{\alpha}%
_{4}\left\langle \tilde{T}^{ij}\tilde{T}^{kl}\right\rangle =2\left( \alpha
_{4}-\beta _{4}\right) \left( \delta ^{il}\delta ^{kj}-\delta ^{ik}\delta
^{lj}\right) \,, \\
\left\langle T^{IJ}Y^{KL}\right\rangle &=&\left\langle \tilde{Y}^{IJ}\tilde{Y%
}^{KL}\right\rangle =\left\langle Y^{IJ}Y^{KL}\right\rangle =\tilde{\alpha}%
_{4}\left\langle \tilde{T}^{IJ}\tilde{T}^{KL}\right\rangle =2\left( \alpha
_{4}+\beta _{4}\right) \left( \delta ^{IL}\delta ^{KJ}-\delta ^{IK}\delta
^{LJ}\right) \,,  \label{pqinvt2}
\end{eqnarray}%
where $\left\{ \tilde{J}_{ab},\tilde{P}_{a},\tilde{T}^{ij},\tilde{T}^{IJ},%
\tilde{Q}_{\alpha }^{i},\tilde{Q}_{\alpha }^{I}\right\} $ correspond to the
original $\mathfrak{osp}\left( 2|p\right) \otimes \mathfrak{osp}\left(
2|q\right) $ generators and where we have defined%
\begin{align*}
\alpha _{0}& \equiv \tilde{\alpha}_{0}\mu _{0}\,,\text{ \ \ \ \ }\alpha
_{2}\equiv \tilde{\alpha}_{2}\mu _{0}\,,\text{ \ \ \ \ }\alpha _{4}\equiv
\tilde{\alpha}_{4}\mu _{0}\,, \\
\beta _{0}& \equiv \tilde{\alpha}_{0}\mu _{1}\,,\text{ \ \ \ \ }\beta
_{2}\equiv \tilde{\alpha}_{2}\mu _{1}\,,\text{ \ \ \ \ }\beta _{4}\equiv
\tilde{\alpha}_{4}\mu _{1}\,.\text{\ }
\end{align*}%
Here $\tilde{\alpha}_{0},\tilde{\alpha}_{2},\tilde{\alpha}_{4}$ are
arbitrary constants as $\mu _{0}$ and $\mu _{1}$. Let us now consider the
gauge connection one-form of this extended superalgebra:

\bigskip
\begin{eqnarray}
A &=&\frac{1}{2}\omega ^{ab}J_{ab}+\frac{1}{2}\tilde{k}^{ab}\tilde{Z}_{ab}+%
\frac{1}{2}k^{ab}Z_{ab}+\frac{1}{l}e^{a}P_{a}+\frac{1}{l}\tilde{h}^{a}\tilde{%
Z}_{a}  \notag \\
&&+\frac{1}{2}A^{ij}T_{ij}+\frac{1}{2}A^{IJ}T_{IJ}+\frac{1}{2}\tilde{B}^{ij}%
\tilde{Y}_{ij}+\frac{1}{2}\tilde{B}^{IJ}\tilde{Y}_{IJ}+\frac{1}{2}%
B^{ij}Y_{ij}\,+\frac{1}{2}B^{IJ}Y_{IJ}  \notag \\
&&+\frac{1}{\sqrt{l}}\bar{\psi}_{i}Q^{i}+\frac{1}{\sqrt{l}}\bar{\psi}%
_{I}Q^{I}+\frac{1}{\sqrt{l}}\bar{\xi}_{i}\Sigma ^{i}+\frac{1}{\sqrt{l}}\bar{%
\xi}_{I}\Sigma ^{I}\,.  \label{pqOF}
\end{eqnarray}%
The curvature two-form $F=dA+AA$ is given by%
\begin{eqnarray}
F &=&\frac{1}{2}R^{ab}J_{ab}+\frac{1}{2}\tilde{F}^{ab}\tilde{Z}_{ab}+\frac{1%
}{2}F^{ab}Z_{ab}+\frac{1}{l}F^{a}P_{a}+\frac{1}{l}\tilde{H}^{a}\tilde{Z}_{a}
\notag \\
&&+\frac{1}{2}F^{ij}T_{ij}+\frac{1}{2}F^{IJ}T_{IJ}+\frac{1}{2}\tilde{G}^{ij}%
\tilde{Y}_{ij}+\frac{1}{2}\tilde{G}^{IJ}\tilde{Y}_{IJ}+\frac{1}{2}%
G^{ij}Y_{ij}+\frac{1}{2}G^{IJ}Y_{IJ}  \notag \\
&&+\frac{1}{\sqrt{l}}\bar{\Psi}_{i}Q^{i}+\frac{1}{\sqrt{l}}\bar{\Psi}%
_{I}Q^{I}+\frac{1}{\sqrt{l}}\bar{\Xi}_{i}\Sigma ^{i}\,+\frac{1}{\sqrt{l}}%
\bar{\Xi}_{I}\Sigma ^{I},
\end{eqnarray}%
where%
\begin{align*}
F^{a}=& D_{\omega }e^{a}+k_{\text{ }b}^{a}e^{b}+\tilde{k}_{\text{ }b}^{a}%
\tilde{h}^{b}-\frac{1}{2}\bar{\psi}^{i}\Gamma ^{a}\psi ^{i}-\frac{1}{2}\bar{%
\xi}^{i}\Gamma ^{a}\xi ^{i}-\frac{1}{2}\bar{\psi}^{I}\Gamma ^{a}\psi ^{I}-%
\frac{1}{2}\bar{\xi}^{I}\Gamma ^{a}\xi ^{I}\,, \\
\tilde{H}^{a}=& D_{\omega }\tilde{h}^{a}+\tilde{k}_{\text{ }b}^{a}e^{b}+k_{%
\text{ }b}^{a}\tilde{h}^{b}-\bar{\psi}^{i}\Gamma ^{a}\xi ^{i}-\bar{\psi}%
^{I}\Gamma ^{a}\xi ^{I}\,, \\
\tilde{F}^{ab}=& D_{\omega }\tilde{k}^{ab}+k_{\text{ }c}^{a}\tilde{k}%
^{cb}-k_{\text{ }c}^{b}\tilde{k}^{ca}+\frac{2}{l^{2}}e^{a}\tilde{h}^{b}+%
\frac{1}{2l}\bar{\psi}^{i}\Gamma ^{ab}\psi ^{i}-\frac{1}{2l}\bar{\psi}%
^{I}\Gamma ^{ab}\psi ^{I} \\
& +\frac{1}{2l}\bar{\xi}^{i}\Gamma ^{ab}\xi ^{i}-\frac{1}{2l}\bar{\xi}%
^{I}\Gamma ^{ab}\xi ^{I}\,, \\
F^{ab}=& D_{\omega }k^{ab}+\tilde{k}_{\text{ }c}^{a}\tilde{k}^{cb}+k_{\text{
}c}^{a}k^{cb}+\frac{1}{l^{2}}e^{a}e^{b}+\frac{1}{l^{2}}\tilde{h}^{a}\tilde{h}%
^{b}+\frac{1}{l}\bar{\xi}^{i}\Gamma ^{ab}\psi ^{i}-\frac{1}{l}\bar{\xi}%
^{I}\Gamma ^{ab}\psi ^{I}\,,
\end{align*}%
\begin{align*}
F^{IJ}=& dA^{IJ}+A^{IK}A^{KJ}\,, \\
\tilde{G}^{IJ}=& d\tilde{B}^{IJ}+A^{IK}\tilde{B}^{KJ}+\tilde{B}^{IK}A^{KJ}+%
\tilde{B}^{IK}B^{KJ}+B^{IK}\tilde{B}^{KJ}-\bar{\psi}^{I}\psi ^{J}\,-\bar{\xi}%
^{I}\xi ^{J}, \\
G^{IJ}=& dB^{IJ}+A^{IK}B^{KJ}+B^{IK}A^{KJ}+\tilde{B}^{IK}\tilde{B}%
^{KJ}+B^{IK}B^{KJ}-2\bar{\psi}^{I}\xi ^{J}\,, \\
\Psi ^{I}=& D_{\omega }\psi ^{I}+\frac{1}{4}k_{ab}\Gamma ^{ab}\psi ^{I}+%
\frac{1}{4}\tilde{k}_{ab}\Gamma ^{ab}\xi ^{I}-\frac{1}{2l}e_{a}\Gamma
^{a}\xi ^{I}-\frac{1}{2l}\tilde{h}_{a}\Gamma ^{a}\psi ^{I} \\
& +A^{IJ}\psi ^{J}+B^{IJ}\psi ^{J}+\tilde{B}^{IJ}\xi ^{J}, \\
\Xi ^{I}=& D_{\omega }\xi ^{I}+\frac{1}{4}k_{ab}\Gamma ^{ab}\xi ^{I}+\frac{1%
}{4}\tilde{k}_{ab}\Gamma ^{ab}\psi ^{I}-\frac{1}{2l}e_{a}\Gamma ^{a}\psi
^{I}-\frac{1}{2l}\tilde{h}_{a}\Gamma ^{a}\xi ^{I}\, \\
& +A^{IJ}\xi ^{J}+B^{IJ}\xi ^{J}+\tilde{B}^{IJ}\psi ^{J}\,,
\end{align*}%
and $R^{ab},F^{ij},\tilde{G}^{ij},G^{ij},\Psi ^{i}$ and $\Sigma ^{i}$ are
defined as in the $\mathcal{N}=2$ case (see Eq. (\ref{N2TF})).

Considering the connection one-form (\ref{pqOF}) and the non-vanishing
components of the invariant tensor ((\ref{invt3a})-(\ref{invt3c}) and (\ref%
{pqinvt})-(\ref{pqinvt2})) in the general three-dimensional CS expression,
we find the CS action of $\mathcal{N}=p+q$ $AdS$-Lorentz supergravity up to
a surface term:

\begin{eqnarray}
I_{CS}^{\left( 2+1\right) } &=&k\int \frac{\alpha _{0}}{2}\left( \omega _{%
\text{ }b}^{a}d\omega _{\text{ }a}^{b}+\frac{2}{3}\omega _{\text{ }%
c}^{a}\omega _{\text{ }b}^{c}\omega _{\text{ }a}^{b}\right) +\left( \alpha
_{0}-\beta _{0}\right) A^{ij}F^{ji}\left( A\right) +\left( \alpha _{0}+\beta
_{0}\right) A^{IJ}F^{JI}\left( A\right)  \notag \\
&&+\frac{\beta _{2}}{l}\epsilon _{abc}\left( R^{ab}e^{c}+\frac{1}{3l^{2}}%
e^{a}e^{b}e^{c}+K^{ab}e^{c}+\tilde{K}^{ab}\tilde{h}^{c}+\frac{1}{l^{2}}%
\tilde{h}^{a}\tilde{h}^{b}e\right)  \notag \\
&&+\left( \alpha _{2}-\beta _{2}\right) \left[ \tilde{B}^{ij}F^{ji}\left(
A\right) +\tilde{B}^{ij}F^{ji}\left( B\right) +\left( A^{ij}+B^{ij}\right)
F^{ji}\left( \tilde{B}\right) \right]  \notag \\
&&+\left( \alpha _{2}+\beta _{2}\right) \left[ \tilde{B}^{IJ}F^{JI}\left(
A\right) +\tilde{B}^{IJ}F^{JI}\left( B\right) +\left( A^{IJ}+B^{IJ}\right)
F^{JI}\left( \tilde{B}\right) \right]  \notag \\
&&+2\left( \beta _{2}-\alpha _{2}\right) \left[ \frac{1}{l}\bar{\psi}%
^{i}\Psi ^{i}+\frac{1}{l}\bar{\xi}^{i}\Xi ^{i}\right] +2\left( \beta
_{2}+\alpha _{2}\right) \left[ \frac{1}{l}\bar{\psi}^{I}\Psi ^{I}+\frac{1}{l}%
\bar{\xi}^{I}\Xi ^{I}\right]  \notag \\
&&+\alpha _{2}\left[ R_{\text{ }b}^{a}\tilde{k}_{\text{ }a}^{b}+K_{\text{ }%
b}^{a}\tilde{k}_{\text{ }a}^{b}+\tilde{K}_{\text{ }b}^{a}k_{\text{ }a}^{b}+%
\frac{1}{l^{2}}e^{a}H_{a}+\frac{1}{l^{2}}\tilde{h}^{a}K_{a}\right]  \notag \\
&&+\frac{\beta _{4}}{l}\epsilon _{abc}\left( R^{ab}\tilde{h}^{c}+\frac{1}{%
3l^{2}}\tilde{h}^{a}\tilde{h}^{b}\tilde{h}^{c}+\tilde{K}^{ab}e^{c}+K^{ab}%
\tilde{h}^{c}+\frac{1}{l^{2}}e^{a}e^{b}\tilde{h}^{c}\right)  \notag \\
&&+\left( \alpha _{4}-\beta _{4}\right) \left[ \tilde{B}^{ij}F^{ji}\left(
\tilde{B}\right) +B^{ij}F^{ji}\left( A\right) +\left( A^{ij}+B^{ij}\right)
F^{ji}\left( B\right) \right]  \notag \\
&&+\left( \alpha _{4}+\beta _{4}\right) \left[ \tilde{B}^{IJ}F^{JI}\left(
\tilde{B}\right) +B^{IJ}F^{JI}\left( A\right) +\left( A^{IJ}+B^{IJ}\right)
F^{JI}\left( B\right) \right]  \notag \\
&&+2\left( \beta _{4}-\alpha _{4}\right) \left[ \frac{1}{l}\bar{\xi}^{i}\Psi
^{i}+\frac{1}{l}\bar{\psi}^{i}\Xi ^{i}\right] +2\left( \beta _{4}+\alpha
_{4}\right) \left[ \frac{1}{l}\bar{\xi}^{I}\Psi ^{I}+\frac{1}{l}\bar{\psi}%
^{I}\Xi ^{I}\right]  \notag \\
&&+\alpha _{4}\left[ R_{\text{ }b}^{a}k_{\text{ }a}^{b}+K_{\text{ }b}^{a}k_{%
\text{ }a}^{b}+\tilde{K}_{\text{ }b}^{a}\tilde{k}_{\text{ }a}^{b}+\frac{1}{%
l^{2}}e^{a}K_{a}+\frac{1}{l^{2}}\tilde{h}^{a}H_{a}\right] \,,
\end{eqnarray}%
where $\Psi ^{i}$, $\Psi ^{I}$, $\Xi ^{i}$ and $\Xi ^{I}$ are the fermionic
components of the curvature two form and%
\begin{eqnarray*}
F^{ij}\left( A\right) &=&dA^{ij}+A^{ik}A^{kj},\text{ \ \ \ \ \ \ \ \ \ \ \ \
\ \ \ \ \ \ }F^{IJ}\left( A\right) =dA^{IJ}+A^{IK}A^{KJ}, \\
F^{ij}\left( \tilde{B}\right) &=&d\tilde{B}^{ij}+A^{ik}\tilde{B}^{kj}+\tilde{%
B}^{ik}A^{kj}+\tilde{B}^{ik}B^{kj}+B^{ik}\tilde{B}^{kj}\,, \\
F^{IJ}\left( \tilde{B}\right) &=&d\tilde{B}^{IJ}+A^{IK}\tilde{B}^{KJ}+\tilde{%
B}^{IK}A^{KJ}+\tilde{B}^{IK}B^{KJ}+B^{IK}\tilde{B}^{KJ}\,, \\
F^{ij}\left( B\right) &=&dB^{ij}+A^{ik}B^{kj}+B^{ik}A^{kj}+\tilde{B}^{ik}%
\tilde{B}^{kj}+B^{ik}B^{kj}\,, \\
F^{IJ}\left( B\right) &=&dB^{IJ}+A^{IK}B^{KJ}+B^{IK}A^{KJ}+\tilde{B}^{IK}%
\tilde{B}^{KJ}+B^{IK}B^{KJ}\,, \\
K^{ab} &=&Dk^{ab}+k_{\text{ }d}^{a}k_{\text{ }b}^{d}+\tilde{k}_{\text{ }%
d}^{a}\tilde{k}_{\text{ }b}^{d}\,,\text{ \ \ \ }\tilde{K}^{ab}=D\tilde{k}%
^{ab}+k_{\text{ }d}^{a}\tilde{k}_{\text{ }b}^{d}+k_{b}^{\text{ }d}\tilde{k}_{%
\text{ }d}^{a}\,, \\
H^{a} &=&D\tilde{h}^{a}+k_{\text{ }b}^{a}\tilde{h}^{b}+\tilde{k}_{\text{ }%
b}^{a}e^{b}\,,\text{ \ \ \ \ \ \ }K^{a}=T^{a}+k_{\text{ }b}^{a}e^{b}+\tilde{k%
}_{\text{ }b}^{a}\tilde{h}^{b}\,.
\end{eqnarray*}

As the $\left( 2,0\right) $ $AdS$-Lorentz superalgebra has its $\left(
2,0\right) $ Maxwell limit, the IW contraction of the $\left( p,q\right) $ $%
AdS$-Lorentz superalgebra leads to the $\left( p,q\right) $ Maxwell
superalgebra. Indeed, by rescaling the $AdS$-Lorentz generators as%
\begin{align*}
\tilde{Z}_{ab}& \rightarrow \sigma ^{2}\tilde{Z}_{ab}\,,\text{ \ \ }%
Z_{ab}\rightarrow \sigma ^{4}Z_{ab}\,,\text{ \ \ }P_{a}\rightarrow \sigma
^{2}P_{a}\,,\text{ \ \ \ \ }J_{ab}\rightarrow J_{ab}\,, \\
\tilde{Z}_{a}& \rightarrow \sigma ^{4}\tilde{Z}_{a}\,,\text{ \ \ }Q_{\alpha
}^{i}\rightarrow \sigma Q_{\alpha }^{i}\,,\text{\ \ }Q_{\alpha
}^{I}\rightarrow \sigma Q_{\alpha }^{I}\,,\text{\ \ }\Sigma ^{i}\rightarrow
\sigma ^{3}\Sigma ^{i}\,,\text{ \ \ } \\
\Sigma ^{I}& \rightarrow \sigma ^{3}\Sigma ^{I}\,,\text{ \ \ }%
Y^{ij}\rightarrow \sigma ^{4}Y^{ij}\,,\text{ \ \ }Y^{IJ}\rightarrow \sigma
^{4}Y^{IJ}\,, \\
\tilde{Y}^{ij}& \rightarrow \sigma ^{2}\tilde{Y}^{ij}\,,\text{ \ \ }\tilde{Y}%
^{IJ}\rightarrow \sigma ^{2}\tilde{Y}^{IJ}\,,\text{ \ \ }T^{ij}\rightarrow
T^{ij}\,,\text{ \ \ }T^{IJ}\rightarrow T^{IJ}\,,
\end{align*}%
and applying the limit $\sigma \rightarrow \infty $, we obtain the $\mathcal{%
N}=p+q$ Maxwell superalgebra whose explicit (anti)commutation relations can
be found in Appendix B. Extending the rescaling of the generators to the\ $%
\alpha $ and $\beta $ constants appearing in the invariant tensor of the $%
\left( p,q\right) $ $AdS$-Lorentz superalgebra as%
\begin{eqnarray}
\alpha _{4} &\rightarrow &\sigma ^{4}\alpha _{4}\,,\text{ \ \ \ \ }\alpha
_{2}\rightarrow \sigma ^{2}\alpha _{2}\,,\text{ \ \ \ \ }\alpha
_{0}\rightarrow \alpha _{0}\,,  \notag \\
\beta _{4} &\rightarrow &\sigma ^{4}\beta _{4}\,,\text{ \ \ \ \ }\,\beta
_{2}\rightarrow \sigma ^{2}\alpha _{1}\,,\text{ \ \ \ \ }\beta
_{0}\rightarrow \beta _{0}\,,  \label{rcts}
\end{eqnarray}%
and considering the limit $\sigma \rightarrow \infty $, we obtain the
non-vanishing components of the invariant tensor for the $\left( p,q\right) $
Maxwell superalgebra:%
\begin{align}
\left\langle J_{ab}J_{cd}\right\rangle _{\mathcal{M}}& =\alpha _{0}\left(
\eta _{ad}\eta _{bc}-\eta _{ac}\eta _{bd}\right) \,,  \label{pqMinvt1} \\
\text{\ \ }\left\langle J_{ab}\tilde{Z}_{cd}\right\rangle _{\mathcal{M}}&
=\alpha _{2}\left( \eta _{ad}\eta _{bc}-\eta _{ac}\eta _{bd}\right) \,, \\
\left\langle \tilde{Z}_{ab}\tilde{Z}_{cd}\right\rangle _{\mathcal{M}}&
=\left\langle J_{ab}Z_{cd}\right\rangle =\alpha _{4}\left( \eta _{ad}\eta
_{bc}-\eta _{ac}\eta _{bd}\right) \,, \\
\left\langle J_{ab}P_{c}\right\rangle _{\mathcal{M}}& =\beta _{2}\epsilon
_{abc}\,,\text{ \ \ \ }\left\langle P_{a}P_{b}\right\rangle _{\mathcal{M}%
}=\alpha _{4}\eta _{ab}\,, \\
\left\langle \tilde{Z}_{ab}P_{c}\right\rangle _{\mathcal{M}}& =\left\langle
J_{ab}\tilde{Z}_{c}\right\rangle _{\mathcal{M}}=\beta _{4}\epsilon _{abc}\,,
\end{align}%
\begin{eqnarray}
\left\langle Q_{\alpha }^{i}Q_{\beta }^{j}\right\rangle _{\mathcal{M}}
&=&2\left( \beta _{2}-\alpha _{2}\right) C_{\alpha \beta }\,\delta ^{ij}\,,%
\text{ \ \ }\left\langle Q_{\alpha }^{I}Q_{\beta }^{J}\right\rangle _{%
\mathcal{M}}=2\left( \beta _{2}+\alpha _{2}\right) C_{\alpha \beta }\,\delta
^{IJ}\,, \\
\left\langle Q_{\alpha }^{i}\Sigma _{\beta }^{j}\right\rangle _{\mathcal{M}}
&=&2\left( \beta _{4}-\alpha _{4}\right) C_{\alpha \beta }\,\delta ^{ij},%
\text{ \ \ \ }\left\langle Q_{\alpha }^{I}\Sigma _{\beta }^{J}\right\rangle
_{\mathcal{M}}=2\left( \beta _{4}+\alpha _{4}\right) C_{\alpha \beta
}\,\delta ^{IJ},
\end{eqnarray}%
\begin{eqnarray}
\left\langle T^{ij}T^{kl}\right\rangle _{\mathcal{M}} &=&2\left( \alpha
_{0}-\beta _{0}\right) \left( \delta ^{il}\delta ^{kj}-\delta ^{ik}\delta
^{lj}\right) \,, \\
\left\langle T^{IJ}T^{KL}\right\rangle _{\mathcal{M}} &=&2\left( \alpha
_{0}+\beta _{0}\right) \left( \delta ^{IL}\delta ^{KJ}-\delta ^{IK}\delta
^{LJ}\right) \,, \\
\left\langle T^{ij}\tilde{Y}^{kl}\right\rangle _{\mathcal{M}} &=&2\left(
\alpha _{2}-\beta _{2}\right) \left( \delta ^{il}\delta ^{kj}-\delta
^{ik}\delta ^{lj}\right) \,, \\
\left\langle T^{IJ}\tilde{Y}^{KL}\right\rangle _{\mathcal{M}} &=&2\left(
\alpha _{2}+\beta _{2}\right) \left( \delta ^{IL}\delta ^{KJ}-\delta
^{IK}\delta ^{LJ}\right) \,, \\
\left\langle T^{ij}Y^{kl}\right\rangle _{\mathcal{M}} &=&\left\langle \tilde{%
Y}^{ij}\tilde{Y}^{kl}\right\rangle _{\mathcal{M}}=2\left( \alpha _{4}-\beta
_{4}\right) \left( \delta ^{il}\delta ^{kj}-\delta ^{ik}\delta ^{lj}\right)
\,, \\
\left\langle T^{IJ}Y^{KL}\right\rangle _{\mathcal{M}} &=&\left\langle \tilde{%
Y}^{IJ}\tilde{Y}^{KL}\right\rangle _{\mathcal{M}}=2\left( \alpha _{4}+\beta
_{4}\right) \left( \delta ^{IL}\delta ^{KJ}-\delta ^{IK}\delta ^{LJ}\right)
\,.  \label{pqMinvt2}
\end{eqnarray}%
Let us note that the generators appearing in the invariant tensor $%
\left\langle \dots \right\rangle _{\mathcal{M}}$ satisfy the $\left(
p,q\right) $ Maxwell superalgebra. Additionally, the rescaling of the
constants considered here preserves the curvatures structure.

Then, the three-dimensional $\left( p,q\right) $ Maxwell supergravity CS
action can be derived considering the $\left( p,q\right) $ Maxwell
connection one-form (analogous to Eq. (\ref{pqOF})) and the non-vanishing
components of the invariant tensor (\ref{pqMinvt1}-\ref{pqMinvt2}):

\begin{eqnarray}
I_{\mathcal{M}-CS}^{\left( 2+1\right) } &=&k\int \frac{\alpha _{0}}{2}\left(
\omega _{\text{ }b}^{a}d\omega _{\text{ }a}^{b}+\frac{2}{3}\omega _{\text{ }%
c}^{a}\omega _{\text{ }b}^{c}\omega _{\text{ }a}^{b}\right) +\frac{\beta _{2}%
}{l}\epsilon _{abc}R^{ab}e^{c}  \notag \\
&&+\left( \alpha _{0}-\beta _{0}\right) A^{ij}F^{ji}\left( A\right) +\left(
\alpha _{0}+\beta _{0}\right) A^{IJ}F^{JI}\left( A\right)  \notag \\
&&+\left( \alpha _{2}-\beta _{2}\right) \left[ \tilde{B}^{ij}F^{ji}\left(
A\right) +A^{ij}F^{ji}\left( \tilde{B}\right) \right] +\left( \alpha
_{2}+\beta _{2}\right) \left[ \tilde{B}^{IJ}F^{JI}\left( A\right)
+A^{IJ}F^{JI}\left( \tilde{B}\right) \right]  \notag \\
&&+2\left( \beta _{2}-\alpha _{2}\right) \frac{1}{l}\bar{\psi}^{i}\Psi
^{i}+2\left( \beta _{2}+\alpha _{2}\right) \frac{1}{l}\bar{\psi}^{I}\Psi ^{I}
\notag \\
&&+\alpha _{2}R_{\text{ }b}^{a}\tilde{k}_{\text{ }a}^{b}+\frac{\beta _{4}}{l}%
\epsilon _{abc}\left( R^{ab}\tilde{h}^{c}+D_{\omega }\tilde{k}%
^{ab}e^{c}\right)  \notag \\
&&+\left( \alpha _{4}-\beta _{4}\right) \left[ \tilde{B}^{ij}F^{ji}\left(
\tilde{B}\right) +B^{ij}F^{ji}\left( A\right) +A^{ij}F^{ji}\left( B\right) %
\right]  \notag \\
&&+\left( \alpha _{4}+\beta _{4}\right) \left[ \tilde{B}^{IJ}F^{JI}\left(
\tilde{B}\right) +B^{IJ}F^{JI}\left( A\right) +A^{IJ}F^{JI}\left( B\right) %
\right]  \notag \\
&&+2\left( \beta _{4}-\alpha _{4}\right) \left[ \frac{1}{l}\bar{\xi}^{i}\Psi
^{i}+\frac{1}{l}\bar{\psi}^{i}\Xi ^{i}\right] +2\left( \beta _{4}+\alpha
_{4}\right) \left[ \frac{1}{l}\bar{\xi}^{I}\Psi ^{I}+\frac{1}{l}\bar{\psi}%
^{I}\Xi ^{I}\right]  \notag \\
&&+\alpha _{4}\left[ R_{\text{ }b}^{a}k_{\text{ }a}^{b}+D_{\omega }\tilde{k}%
_{\text{ }b}^{a}\tilde{k}_{\text{ }a}^{b}+\frac{1}{l^{2}}e^{a}T_{a}\right]
\,,  \label{pqMCS}
\end{eqnarray}%
where $\Psi ^{i}$, $\Psi ^{I}$, $\Xi ^{i}$ and $\Xi ^{I}$ are the fermionic
components of the $\left( p,q\right) $ Maxwell curvature two form, given by%
\begin{eqnarray*}
\Psi ^{i} &=&d\psi ^{i}+\frac{1}{4}\omega _{ab}\Gamma ^{ab}\psi
^{i}+A^{ij}\psi ^{j}\,, \\
\Psi ^{I} &=&d\psi ^{I}+\frac{1}{4}\omega _{ab}\Gamma ^{ab}\psi
^{I}+A^{IJ}\psi ^{J}\,, \\
\Xi ^{i} &=&d\xi ^{i}+\frac{1}{4}\omega _{ab}\Gamma ^{ab}\xi ^{i}+\frac{1}{4}%
\tilde{k}_{ab}\Gamma ^{ab}\psi ^{i}+\frac{1}{2l}e_{a}\Gamma ^{a}\psi
^{i}+A^{ij}\xi ^{j}+\tilde{B}^{ij}\psi ^{j}\,, \\
\Xi ^{I} &=&d\xi ^{I}+\frac{1}{4}\omega _{ab}\Gamma ^{ab}\xi ^{I}+\frac{1}{4}%
\tilde{k}_{ab}\Gamma ^{ab}\psi ^{I}-\frac{1}{2l}e_{a}\Gamma ^{a}\psi
^{I}+A^{IJ}\xi ^{J}+\tilde{B}^{IJ}\psi ^{J}\,\,,
\end{eqnarray*}%
and%
\begin{eqnarray*}
F^{ij}\left( A\right) &=&dA^{ij}+A^{ik}A^{kj},\text{ \ \ \ \ }F^{IJ}\left(
A\right) =dA^{IJ}+A^{IK}A^{KJ}, \\
F^{ij}\left( \tilde{B}\right) &=&d\tilde{B}^{ij}+A^{ik}\tilde{B}^{kj}+\tilde{%
B}^{ik}A^{kj}\,, \\
F^{IJ}\left( \tilde{B}\right) &=&d\tilde{B}^{IJ}+A^{IK}\tilde{B}^{KJ}+\tilde{%
B}^{IK}A^{KJ}\,, \\
F^{ij}\left( B\right) &=&dB^{ij}+A^{ik}B^{kj}+B^{ik}A^{kj}+\tilde{B}^{ik}%
\tilde{B}^{kj}, \\
F^{IJ}\left( B\right) &=&dB^{IJ}+A^{IK}B^{KJ}+B^{IK}A^{KJ}+\tilde{B}^{IK}%
\tilde{B}^{KJ}\,.
\end{eqnarray*}%
One can note that the $\left( p,q\right) $ Maxwell supergravity action can
be obtained directly from the $AdS$-Lorentz one considering the rescaling of
the constant (given by Eq. (\ref{rcts}) ) and the gauge fields%
\begin{eqnarray*}
\omega _{ab} &\rightarrow &\omega _{ab}\,,\text{ \ \ \ }\tilde{k}%
_{ab}\rightarrow \sigma ^{-2}\tilde{k}_{ab}\,,\text{ \ \ \ }%
k_{ab}\rightarrow \sigma ^{-4}k_{ab}\,,\text{ \ \ \ }e_{a}\rightarrow \sigma
^{-2}e_{a}\,,\text{ \ \ \ \ }\tilde{h}_{a}\rightarrow \sigma ^{-4}\tilde{h}%
_{a}\,, \\
\psi ^{i} &\rightarrow &\sigma ^{-1}\psi ^{i}\,,\text{ \ \ \ \ }\psi
^{I}\rightarrow \sigma ^{-1}\psi ^{I}\,,\text{ \ \ \ }\xi ^{i}\rightarrow
\sigma ^{-3}\xi ^{i}\,,\text{ \ \ \ }\xi ^{I}\rightarrow \sigma ^{-3}\xi
^{I}\,,\text{ \ \ \ }A^{ij}\rightarrow A^{ij}\,,\text{ \ \ \ } \\
A^{IJ} &\rightarrow &A^{IJ}\,,\text{ \ \ \ }\tilde{B}^{ij}\rightarrow \sigma
^{-2}\tilde{B}^{ij}\,,\text{ \ \ \ }\tilde{B}^{IJ}\rightarrow \sigma ^{-2}%
\tilde{B}^{IJ}\,,\text{ \ \ \ }B^{ij}\rightarrow \sigma ^{-4}B^{ij}\,,\text{
\ \ \ }B^{IJ}\rightarrow \sigma ^{-4}B^{IJ}\,,
\end{eqnarray*}%
and then the limit $\sigma \rightarrow \infty $. Thus the procedure
presented here assures the explicit Maxwell limit considering the rescaling
not only of the fields but also of the constants appearing in the invariant
tensor. Naturally, the Poincar\'{e} limit approach presented in Ref.~\cite%
{HIPT} could be applied here, but it would require the introduction of
additional\ gauge fields. This would lead not only to new Maxwell
supergravity theories but also to new $AdS$-Lorentz ones.

\section{Conclusion}

We have presented a generalization of the standard In\"{o}n\"{u}-Wigner
contraction combining the rescaling of the generators and the constants
appearing in the invariant tensor of a Lie superalgebra. The procedure
considered here allows not only to obtain the invariant tensor of a
contracted superalgebra but also to construct the contracted supergravity
action.

In particular, \ we have shown that the Poincar\'{e} limit can be applied to
a $\left( p,q\right) $ $AdS$ supergravity in presence of the exotic
Lagrangian without introducing an $\mathfrak{so}\left( p\right) \oplus
\mathfrak{so}\left( q\right) $ extension. Naturally, the internal symmetries
generators of the $AdS$ superalgebra behave as central charges after the
contraction and do not contribute explicitly in the construction of the
Poincar\'{e} action. Additionally, no gravitino fields contribute to the
exotic form. It is important to point out that the standard IW contraction
does not allow to apply the Poincar\'{e} limit to the $\left( p,q\right) $ $%
AdS$ supergravity in presence of the Pontryagin-Chern-Simons form.

We have also applied our generalized IW scheme to expanded superalgebras. We
have constructed a new class of $D=2+1$ $\left( p,q\right) $ Maxwell
supergravity theories as a particular limit of an $AdS$-Lorentz supergravity
model. The results presented here show an explicit relation between
contraction and expansion. Besides, we have shown that the fermionic and the
internal symmetries fields contribute to the exotic CS form. Nevertheless,
we have clarified that a supergravity with Maxwell supersymmetry requires
the introduction of an additional spinorial field $\xi $.

The procedure considered here could be useful in higher-dimensional
supergravity models in order to derive non trivial Chern-Simons supergravity
theories (work in progress). It would be interesting to explore the
expansion and contraction method in the context of infinite dimensional
algebras and hypergravity.

It would be also interesting to extend our study of the Maxwell
supergravities to black hole solutions. It is of particular interest to
study black hole solutions with torsion for their thermodynamical properties
\cite{BC1, BC2, TZ}. In particular, one could explore the possibility of
finding Maxwell "exotic" BTZ type black holes.

\section{Acknowledgment}

This work was supported by the Newton-Picarte CONICYT Grant No. DPI20140053.
(P.K.C. and E.K.R.) and by the Conicyt - PAI grant No. 79150061 (P.K.C.).

{\small \appendix}

\section{Appendix\label{App}}

The $\mathcal{N}$-extended $AdS$-Lorentz superalgebra is generated by the
set of generators
\begin{equation*}
\left\{ J_{ab},P_{a},\tilde{Z}_{ab},\tilde{Z}_{a},Z_{ab},T^{ij},\tilde{Y}%
^{ij},Y^{ij},Q_{\alpha }^{i},\Sigma _{\alpha }^{i}\right\}
\end{equation*}%
(with $i=1,\dots ,\mathcal{N};$ $\alpha =1,\dots ,4$) whose generators
satisfy the (anti)-commutation relations:%
\begin{align}
\left[ J_{ab},J_{cd}\right] & =\eta _{bc}J_{ad}-\eta _{ac}J_{bd}-\eta
_{bd}J_{ac}+\eta _{ad}J_{bc}\,, \\
\left[ Z_{ab},Z_{cd}\right] & =\eta _{bc}Z_{ad}-\eta _{ac}Z_{bd}-\eta
_{bd}Z_{ac}+\eta _{ad}Z_{bc}\,, \\
\left[ J_{ab},Z_{cd}\right] & =\eta _{bc}Z_{ad}-\eta _{ac}Z_{bd}-\eta
_{bd}Z_{ac}+\eta _{ad}Z_{bc}\,, \\
\left[ J_{ab},\tilde{Z}_{cd}\right] & =\eta _{bc}\tilde{Z}_{ad}-\eta _{ac}%
\tilde{Z}_{bd}-\eta _{bd}\tilde{Z}_{ac}+\eta _{ad}\tilde{Z}_{bc}\,, \\
\left[ \tilde{Z}_{ab},\tilde{Z}_{cd}\right] & =\eta _{bc}Z_{ad}-\eta
_{ac}Z_{bd}-\eta _{bd}Z_{ac}+\eta _{ad}Z_{bc}\,, \\
\left[ \tilde{Z}_{ab},Z_{cd}\right] & =\eta _{bc}\tilde{Z}_{ad}-\eta _{ac}%
\tilde{Z}_{bd}-\eta _{bd}\tilde{Z}_{ac}+\eta _{ad}\tilde{Z}_{bc}\,,
\end{align}%
\begin{align}
\left[ J_{ab},P_{c}\right] & =\eta _{bc}P_{a}-\eta _{ac}P_{b}\,,\text{ \ \ \
\ }\left[ Z_{ab},P_{c}\right] =\eta _{bc}P_{a}-\eta _{ac}P_{b}\,,\text{ } \\
\left[ \tilde{Z}_{ab},P_{c}\right] & =\eta _{bc}\tilde{Z}_{a}-\eta _{ac}%
\tilde{Z}_{b}\,,\text{ \ \ \ \ }\left[ J_{ab},\tilde{Z}_{c}\right] =\eta
_{bc}\tilde{Z}_{a}-\eta _{ac}\tilde{Z}_{b}\,, \\
\left[ \tilde{Z}_{ab},\tilde{Z}_{c}\right] & =\eta _{bc}P_{a}-\eta
_{ac}P_{b}\,,\text{ \ \ \ \ }\left[ Z_{ab},\tilde{Z}_{c}\right] =\eta _{bc}%
\tilde{Z}_{a}-\eta _{ac}\tilde{Z}_{b}\,,  \label{dif1} \\
\left[ P_{a},P_{b}\right] & =Z_{ab}\,,\text{ \ \ \ \ }\left[ \tilde{Z}%
_{a},P_{b}\right] =\tilde{Z}_{ab}\,,\text{ \ \ \ \ }\left[ \tilde{Z}_{a},%
\tilde{Z}_{b}\right] =Z_{ab}\,,  \label{dif2}
\end{align}%
\begin{align}
\left[ T^{ij},T^{kl}\right] & =\delta ^{jk}T^{il}-\delta ^{ik}T^{jl}-\delta
^{jl}T^{ik}+\delta ^{il}T^{jk}\,, \\
\left[ T^{ij},Y^{kl}\right] & =\delta ^{jk}Y^{il}-\delta ^{ik}Y^{jl}-\delta
^{jl}Y^{ik}+\delta ^{il}Y^{jk}\,, \\
\left[ T^{ij},\tilde{Y}^{kl}\right] & =\delta ^{jk}\tilde{Y}^{il}-\delta
^{ik}\tilde{Y}^{jl}-\delta ^{jl}\tilde{Y}^{ik}+\delta ^{il}\tilde{Y}^{jk}\,,
\\
\left[ \tilde{Y}^{ij},\tilde{Y}^{kl}\right] & =\delta ^{jk}Y^{il}-\delta
^{ik}Y^{jl}-\delta ^{jl}Y^{ik}+\delta ^{il}Y^{jk}\,, \\
\left[ \tilde{Y}^{ij},Y^{kl}\right] & =\delta ^{jk}\tilde{Y}^{il}-\delta
^{ik}\tilde{Y}^{jl}-\delta ^{jl}\tilde{Y}^{ik}+\delta ^{il}\tilde{Y}^{jk}\,,
\\
\left[ Y^{ij},Y^{kl}\right] & =\delta ^{jk}Y^{il}-\delta ^{ik}Y^{jl}-\delta
^{jl}Y^{ik}+\delta ^{il}Y^{jk}\,,
\end{align}%
\begin{align}
\left[ J_{ab},Q_{\alpha }^{i}\right] & =-\frac{1}{2}\left( \Gamma
_{ab}Q^{i}\right) _{\alpha }\,,\text{ \ \ \ \ }\left[ P_{a},Q_{\alpha }^{i}%
\right] =-\frac{1}{2}\left( \Gamma _{a}\Sigma ^{i}\right) _{\alpha }\,, \\
\left[ \tilde{Z}_{ab},Q_{\alpha }^{i}\right] & =-\frac{1}{2}\left( \Gamma
_{ab}\Sigma ^{i}\right) _{\alpha }\,,\text{ \ \ \ \ }\left[ \tilde{Z}%
_{a},Q_{\alpha }^{i}\right] =-\frac{1}{2}\left( \Gamma _{a}Q^{i}\right)
_{\alpha }\,, \\
\left[ Z_{ab},Q_{\alpha }^{i}\right] & =-\frac{1}{2}\left( \Gamma
_{ab}Q^{i}\right) _{\alpha }\,,\text{ \ \ \ \ }\left[ P_{a},\Sigma _{\alpha
}^{i}\right] =-\frac{1}{2}\left( \Gamma _{a}Q^{i}\right) _{\alpha }\,, \\
\left[ J_{ab},\Sigma _{\alpha }^{i}\right] & =-\frac{1}{2}\left( \Gamma
_{ab}\Sigma ^{i}\right) _{\alpha }\,,\text{ \ \ \ \ }\left[ \tilde{Z}%
_{a},\Sigma _{\alpha }^{i}\right] =-\frac{1}{2}\left( \Gamma _{a}\Sigma
^{i}\right) _{\alpha }\,, \\
\left[ \tilde{Z}_{ab},\Sigma _{\alpha }^{i}\right] & =-\frac{1}{2}\left(
\Gamma _{ab}Q^{i}\right) _{\alpha }\,,\text{ \ \ \ \ }\left[ Z_{ab},\Sigma
_{\alpha }^{i}\right] =-\frac{1}{2}\left( \Gamma _{ab}\Sigma ^{i}\right)
_{\alpha }\,, \\
\left[ T^{ij},Q_{\alpha }^{k}\right] & =(\delta ^{jk}Q_{\alpha }^{i}-\delta
^{ik}Q_{\alpha }^{j})\,,\text{ \ \ }\left[ \tilde{Y}^{ij},Q_{\alpha }^{k}%
\right] =(\delta ^{jk}\Sigma _{\alpha }^{i}-\delta ^{ik}\Sigma _{\alpha
}^{j})\,, \\
\left[ Y^{ij},Q_{\alpha }^{k}\right] & =(\delta ^{jk}Q_{\alpha }^{i}-\delta
^{ik}Q_{\alpha }^{j})\,,\text{ \ \ }\left[ T^{ij},\Sigma _{\alpha }^{k}%
\right] =(\delta ^{jk}\Sigma _{\alpha }^{i}-\delta ^{ik}\Sigma _{\alpha
}^{j})\,, \\
\left[ \tilde{Y}^{ij},\Sigma _{\alpha }^{k}\right] & =(\delta ^{jk}Q_{\alpha
}^{i}-\delta ^{ik}Q_{\alpha }^{j})\,,\text{ \ \ }\left[ Y^{ij},\Sigma
_{\alpha }^{k}\right] =(\delta ^{jk}\Sigma _{\alpha }^{i}-\delta ^{ik}\Sigma
_{\alpha }^{j})\,,
\end{align}%
\begin{align}
\left\{ Q_{\alpha }^{i},Q_{\beta }^{j}\right\} & =-\frac{1}{2}\delta ^{ij}%
\left[ \left( \Gamma ^{ab}C\right) _{\alpha \beta }\tilde{Z}_{ab}-2\left(
\Gamma ^{a}C\right) _{\alpha \beta }P_{a}\right] \,+C_{\alpha \beta }\tilde{Y%
}^{ij}, \\
\left\{ Q_{\alpha }^{i},\Sigma _{\beta }^{j}\right\} & =-\frac{1}{2}\delta
^{ij}\left[ \left( \Gamma ^{ab}C\right) _{\alpha \beta }Z_{ab}-2\left(
\Gamma ^{a}C\right) _{\alpha \beta }\tilde{Z}_{a}\right] +C_{\alpha \beta
}Y^{ij}\,, \\
\left\{ \Sigma _{\alpha }^{i},\Sigma _{\beta }^{j}\right\} & =-\frac{1}{2}%
\delta ^{ij}\left[ \left( \Gamma ^{ab}C\right) _{\alpha \beta }\tilde{Z}%
_{ab}-2\left( \Gamma ^{a}C\right) _{\alpha \beta }P_{a}\right] \,+C_{\alpha
\beta }\tilde{Y}^{ij}.
\end{align}%
Let us note that the $\mathcal{N}=1$ case (whose internal symmetries
generators $T^{ij}$, $Y^{ij}$ and $\tilde{Y}^{ij}$ are absents) reproduces
the minimal $AdS$-Lorentz superalgebra introduced in Ref.~\cite{CRS}.

\section{Appendix\label{App2}}

The $\left( p,q\right) $ Maxwell superalgebra is generated by the following
set of%
\begin{equation*}
\left\{ J_{ab},P_{a},\tilde{Z}_{ab},\tilde{Z}_{a},Z_{ab},T^{ij},T^{IJ},%
\tilde{Y}^{ij},\tilde{Y}^{IJ},Y^{ij},Y^{IJ},Q_{\alpha }^{i},Q_{\alpha
}^{I},\Sigma _{\alpha }^{i},\Sigma _{\alpha }^{I}\right\}
\end{equation*}%
with $i=1,\dots ,p$ and $I=1,\dots ,q$. The $\left( p,q\right) $ Maxwell
generators satisfy the following (anti)commutation relations%
\begin{align}
\left[ J_{ab},J_{cd}\right] & =\eta _{bc}J_{ad}-\eta _{ac}J_{bd}-\eta
_{bd}J_{ac}+\eta _{ad}J_{bc}\,, \\
\left[ J_{ab},Z_{cd}\right] & =\eta _{bc}Z_{ad}-\eta _{ac}Z_{bd}-\eta
_{bd}Z_{ac}+\eta _{ad}Z_{bc}\,, \\
\left[ J_{ab},\tilde{Z}_{cd}\right] & =\eta _{bc}\tilde{Z}_{ad}-\eta _{ac}%
\tilde{Z}_{bd}-\eta _{bd}\tilde{Z}_{ac}+\eta _{ad}\tilde{Z}_{bc}\,, \\
\left[ \tilde{Z}_{ab},\tilde{Z}_{cd}\right] & =\eta _{bc}Z_{ad}-\eta
_{ac}Z_{bd}-\eta _{bd}Z_{ac}+\eta _{ad}Z_{bc}\,, \\
\left[ J_{ab},P_{c}\right] & =\eta _{bc}P_{a}-\eta _{ac}P_{b}\,,\text{ \ \ \
\ }\left[ P_{a},P_{b}\right] =Z_{ab}\,, \\
\left[ \tilde{Z}_{ab},P_{c}\right] & =\eta _{bc}\tilde{Z}_{a}-\eta _{ac}%
\tilde{Z}_{b}\,,\text{ \ \ \ \ }\left[ J_{ab},\tilde{Z}_{c}\right] =\eta
_{bc}\tilde{Z}_{a}-\eta _{ac}\tilde{Z}_{b}\,,
\end{align}%
\begin{align}
\left[ T^{ij},T^{kl}\right] & =\delta ^{jk}T^{il}-\delta ^{ik}T^{jl}-\delta
^{jl}T^{ik}+\delta ^{il}T^{jk}\,, \\
\left[ T^{ij},Y^{kl}\right] & =\delta ^{jk}Y^{il}-\delta ^{ik}Y^{jl}-\delta
^{jl}Y^{ik}+\delta ^{il}Y^{jk}\,, \\
\left[ T^{ij},\tilde{Y}^{kl}\right] & =\delta ^{jk}\tilde{Y}^{il}-\delta
^{ik}\tilde{Y}^{jl}-\delta ^{jl}\tilde{Y}^{ik}+\delta ^{il}\tilde{Y}^{jk}\,,
\\
\left[ \tilde{Y}^{ij},\tilde{Y}^{kl}\right] & =\delta ^{jk}Y^{il}-\delta
^{ik}Y^{jl}-\delta ^{jl}Y^{ik}+\delta ^{il}Y^{jk}\,,
\end{align}%
\begin{eqnarray}
\left[ T^{IJ},T^{KL}\right] &=&\delta ^{JK}T^{IL}-\delta ^{IK}T^{JL}-\delta
^{JL}T^{IK}+\delta ^{IL}T^{JK}\text{\thinspace }, \\
\left[ T^{IJ},\tilde{Y}^{KL}\right] &=&\delta ^{JK}\tilde{Y}^{IL}-\delta
^{IK}\tilde{Y}^{JL}-\delta ^{JL}\tilde{Y}^{IK}+\delta ^{IL}\tilde{Y}^{JK}%
\text{\thinspace }, \\
\left[ T^{IJ},Y^{KL}\right] &=&\delta ^{JK}Y^{IL}-\delta ^{IK}Y^{JL}-\delta
^{JL}Y^{IK}+\delta ^{IL}Y^{JK}\text{\thinspace }, \\
\left[ \tilde{Y}^{IJ},\tilde{Y}^{KL}\right] &=&\delta ^{JK}Y^{IL}-\delta
^{IK}Y^{JL}-\delta ^{JL}Y^{IK}+\delta ^{IL}Y^{JK}\text{\thinspace },
\end{eqnarray}%
\begin{align}
\left[ J_{ab},Q_{\alpha }^{i}\right] & =-\frac{1}{2}\left( \Gamma
_{ab}Q^{i}\right) _{\alpha }\,,\text{ \ \ \ \ }\left[ P_{a},Q_{\alpha }^{i}%
\right] =-\frac{1}{2}\left( \Gamma _{a}\Sigma ^{i}\right) _{\alpha }\,, \\
\left[ \tilde{Z}_{ab},Q_{\alpha }^{i}\right] & =-\frac{1}{2}\left( \Gamma
_{ab}\Sigma ^{i}\right) _{\alpha }\,,\text{ \ \ \ \ }\left[ J_{ab},\Sigma
_{\alpha }^{i}\right] =-\frac{1}{2}\left( \Gamma _{ab}\Sigma ^{i}\right)
_{\alpha }\,, \\
\left[ T^{ij},Q_{\alpha }^{k}\right] & =(\delta ^{jk}Q_{\alpha }^{i}-\delta
^{ik}Q_{\alpha }^{j})\,,\text{ \ \ }\left[ \tilde{Y}^{ij},Q_{\alpha }^{k}%
\right] =(\delta ^{jk}\Sigma _{\alpha }^{i}-\delta ^{ik}\Sigma _{\alpha
}^{j})\,, \\
\left[ T^{ij},\Sigma _{\alpha }^{k}\right] & =(\delta ^{jk}\Sigma _{\alpha
}^{i}-\delta ^{ik}\Sigma _{\alpha }^{j})\,, \\
\left[ J_{ab},Q_{\alpha }^{I}\right] & =-\frac{1}{2}\left( \Gamma
_{ab}Q^{I}\right) _{\alpha }\,,\text{ \ \ \ \ }\left[ P_{a},Q_{\alpha }^{I}%
\right] =\frac{1}{2}\left( \Gamma _{a}\Sigma ^{I}\right) _{\alpha }\,, \\
\left[ \tilde{Z}_{ab},Q_{\alpha }^{I}\right] & =-\frac{1}{2}\left( \Gamma
_{ab}\Sigma ^{I}\right) _{\alpha }\,,\text{ \ \ \ \ }\left[ J_{ab},\Sigma
_{\alpha }^{I}\right] =-\frac{1}{2}\left( \Gamma _{ab}\Sigma ^{I}\right)
_{\alpha }\,, \\
\left[ T^{IJ},Q_{\alpha }^{K}\right] & =(\delta ^{JK}Q_{\alpha }^{I}-\delta
^{IK}Q_{\alpha }^{J})\,,\text{ \ \ }\left[ \tilde{Y}^{IJ},Q_{\alpha }^{K}%
\right] =(\delta ^{JK}\Sigma _{\alpha }^{I}-\delta ^{IK}\Sigma _{\alpha
}^{J})\,, \\
\left[ T^{IJ},\Sigma _{\alpha }^{K}\right] & =(\delta ^{JK}\Sigma _{\alpha
}^{I}-\delta ^{IK}\Sigma _{\alpha }^{J})\,,
\end{align}%
\begin{align}
\left\{ Q_{\alpha }^{i},Q_{\beta }^{j}\right\} & =-\frac{1}{2}\delta ^{ij}%
\left[ \left( \Gamma ^{ab}C\right) _{\alpha \beta }\tilde{Z}_{ab}-2\left(
\Gamma ^{a}C\right) _{\alpha \beta }P_{a}\right] \,+C_{\alpha \beta }\tilde{Y%
}^{ij}, \\
\left\{ Q_{\alpha }^{i},\Sigma _{\beta }^{j}\right\} & =-\frac{1}{2}\delta
^{ij}\left[ \left( \Gamma ^{ab}C\right) _{\alpha \beta }Z_{ab}-2\left(
\Gamma ^{a}C\right) _{\alpha \beta }\tilde{Z}_{a}\right] +C_{\alpha \beta
}Y^{ij}\,, \\
\left\{ Q_{\alpha }^{I},Q_{\beta }^{J}\right\} & =\frac{1}{2}\delta ^{IJ}%
\left[ \left( \Gamma ^{ab}C\right) _{\alpha \beta }\tilde{Z}_{ab}+2\left(
\Gamma ^{a}C\right) _{\alpha \beta }P_{a}\right] \,-C_{\alpha \beta }\tilde{Y%
}^{IJ}, \\
\left\{ Q_{\alpha }^{I},\Sigma _{\beta }^{J}\right\} & =\frac{1}{2}\delta
^{IJ}\left[ \left( \Gamma ^{ab}C\right) _{\alpha \beta }Z_{ab}+2\left(
\Gamma ^{a}C\right) _{\alpha \beta }\tilde{Z}_{a}\right] -C_{\alpha \beta
}Y^{IJ}\,.
\end{align}%
One can note that when $q=0$, we recover the usual $\mathcal{N}$-extended
Maxwell superalgebra whose (anti)commutation relations can be found in Refs.~%
\cite{AILW, CR1}.


\begin{thebibliography}{99}
\bibitem{AT} A. Achucarro, P.K. Townsend, \textit{A Chern-Simons Action For
Three-Dimensional Anti-De Sitter Supergravity Theories}, Phys. Lett. B
\textbf{180} (1986) 89.

\bibitem{W} E. Witten, \textit{(2+1)-Dimensional gravity as an exactly
soluble system}, Nucl. Phys. B \textbf{311} (1988) 46.

\bibitem{DK} S. Deser, J.H. Kay, \textit{Topologically Massive Supergravity}%
, Phys. Lett. B \textbf{120} (1983) 97.

\bibitem{D} S. Deser, \textit{Cosmological Topological Supergravity},
Quantum Theory of Gravity: Essays in honor of the 60th Birthday of Bryce S.
DeWitt. Published by Adam Hilger Ltd., Bristol, England 1984, p.374.

\bibitem{PvN1} P. van Nieuwenhuizen, \textit{Three-dimensional conformal
supergravity and Chern-Simons terms}, Phys. Rev. D \textbf{32} (1985) 872.

\bibitem{RPvN} M. Rocek, P. van Nieuwenhuizen,\textit{\ }$N\geq 2$\textit{\
supersymmetric Chern-Simons terms as }$d=3$\textit{\ extended conformal
supergravity}, Class. Quant. Grav. \textbf{3} (1986) 43.

\bibitem{AT2} A. Achucarro, P.K. Townsend, \textit{Extended supergravity in
d=(2+1) as Chern-Simons theories}, Phys. Lett. B \textbf{229} (1989) 383.

\bibitem{NG} H. Nishino, S.J. Gates Jr., \textit{Chern-Simons theories with
supersymmetries in three-dimensions}, Mod. Phys. A \textbf{8} (1993) 3371.

\bibitem{HIPT} P.S. Howe, J.M. Izquierdo, G. Papadopoulos, P.K. Townsend,
\textit{New supergravities with central charges and Killing spinors in 2+1
dimensions}, Nucl. Phys. B \textbf{467} (1996) 183. [hep-th/9505032].

\bibitem{BTZ} M. Banados, R. Troncoso, J. Zanelli, \textit{Higher
dimensional Chern-Simons supergravity}, Phys. Rev. D \textbf{54} (1996)
2605. [gr-qc/9601003].

\bibitem{GTW} A. Giacomini, R. Troncoso, S. Willison, \textit{%
Three-dimensional supergravity reloaded}, Class. Quant. Grav. \textbf{24}
(2007) 2845. [hep-th/0610077].

\bibitem{GS} R.K. Gupta, A. Sen, \textit{Consistent Truncation to Three
Dimensional (Super-)gravity}, JHEP \textbf{0803} (2008) 015. arXiv:0710.4177
[hep-th].

\bibitem{ABRHST} R. Andringa, E.A. Bergshoeff, M. de Roo, O. Hohm, E.
Sezgin, P.K. Townsend, \textit{Massive 3D Supergravity}, Class. Quant. Grav.
\textbf{27} (2010) 025010. arXiv:0907.4658 [hep-th].

\bibitem{BHRT} E.A. Bergshoeff, O. Hohm, J. Rosseel, P.K. Townsend, \textit{%
On Maximal Massive 3D Supergravity}, Class. Quant. Grav. \textbf{27} (2010)
235012. arXiv:1007.4075 [hep-th].

\bibitem{BKPRYZ} E.A. Bergshoeff, M. Kovacevic, L. Parra, J. Rosseel, Y.
Yin, T. Zojer, \textit{New Massive Supergravity and Auxiliary Fields},
Class. Quant. Grav. \textbf{30} (2013) 195004. arXiv:1304.5445 [hep-th].

\bibitem{ABRS} R. Andringa, E.A. Bergshoeff, J. Rosseel, E. Sezgin, \textit{%
3D Newton-Cartan supergravity}, Class. Quant. Grav. \textbf{30} (2013)
205005. arXiv:1305.6737 [hep-th].

\bibitem{BKNTM} D. Butter, S. M. Kuzenko, J. Novak, G.
Tartaglino-Mazzucchelli, \textit{Conformal supegravity in three dimensions:
Off-shell actions}, JHEP \textbf{1310} (2013) 073. arXiv:1306.1205 [hep-th].

\bibitem{NT} M. Nishimura, Y. Tanii, \textit{N=6 conformal supergravity in
three dimensions}, JHEP \textbf{1310} (2013) 123. arXiv:1308.3960 [hep-th].

\bibitem{FISV} O. Fierro, F. Izaurieta, P. Salgado, O. Valdivia, \textit{%
(2+1)-dimensional supergravity invariant under the AdS-Lorentz superalgebra}%
, arXiv:1401.3697 [hep-th].

\bibitem{ABBOS} G. Alkac, L. Basanisi, E.A. Bergshoeff, M. Ozkan, E. Sezgin,
\textit{Massive N=2 Supergravity in Three Dimensions}, JHEP \textbf{1502}
(2015) 125. arXiv:1412.3118 [hep-th].

\bibitem{FMT} O. Fuentealba, J. Matulich, R. Troncoso, \textit{Extension of
the Poincar\'{e} group with half-integer spin generators: hypergravity and
beyond}, JHEP \textbf{1509} (2015) 003. arXiv:1505.06173 [hep-th].

\bibitem{CFRS} P.K. Concha, O. Fierro, E.K. Rodr\'{\i}guez, P. Salgado,
\textit{Chern-Simons Supergravity in D=3 and Maxwell superalgebra}, Phys.
Lett. B \textbf{750} (2015) 117. arXiv:1507.02335 [hep-th].

\bibitem{FMT2} O. Fuentealba, J. Matulich, R. Troncoso, \textit{%
Asymptotically flat structure of hypergravity in three spacetime dimensions}%
, JHEP \textbf{1510} (2015) 009. arXiv:1508.04663 [hep-th].

\bibitem{BRZ} E.A. Bergshoeff, J. Rosseel, T. Zojer, \textit{Newton-Cartan
supergravity with torsion and Schr\"{o}dinger supergravity}, JHEP \textbf{%
1511} (2015) 180. arXiv:1509.04527 [hep-th].

\bibitem{BDMT} G. Barnich, L. Donnay, J. Matulich, R. Troncoso,\textit{\
Super-BMS}$_{3}$\textit{\ invariant boundary theory from three-dimensional
flat supergravity}, arXiv:1510.08824 [hep-th].

\bibitem{KRR} C. Krishnan, A. Raju, S. Roy, \textit{A Grassmann path from }$%
AdS_{3}$ \textit{to flat space}, JHEP \textbf{1403} (2014) 036.
arXiv:1312.2941 [hep-th].

\bibitem{LM} I. Lodato, W. Merbis, \textit{Super-BMS}$_{3}$\textit{\
algebras from N=2 flat supergravities}, arXiv:1610.07506 [hep-th].

\bibitem{GRCS} F. Izaurieta, P. Minning, A. Perez, E. Rodr\'{\i}guez, P.
Salgado, \textit{Standard General Relativity from Chern-Simons Gravity},%
\textit{\ }Phys. Lett. B \textbf{678} (2009) 213. arXiv:0905.2187 [hep-th].

\bibitem{CPRS1} P.K. Concha, D.M. Pe\~{n}afiel, E.K. Rodr\'{\i}guez, P.
Salgado, \textit{Even-dimensional General Relativity from Born-Infeld gravity%
}, Phys. Lett. B \textbf{725} (2013) 419. arXiv:1309.0062 [hep-th].

\bibitem{CPRS2} P.K. Concha, D.M. Pe\~{n}afiel, E.K. Rodr\'{\i}guez, P.
Salgado, \textit{Chern-Simons and Born-Infeld gravity theories and Maxwell
algebras type}, Eur. Phys. J. C \textbf{74} (2014) 2741. arXiv:1402.0023
[hep-th].

\bibitem{CPRS3} P.K. Concha, D.M. Pe\~{n}afiel, E.K. Rodr\'{\i}guez, P.
Salgado, \textit{Generalized Poincare algebras and Lovelock-Cartan gravity
theory}, Phys. Lett. B \textbf{742} (2015) 310. arXiv:1405.7078 [hep-th].

\bibitem{Cai:2006pq} R.~G.~Cai and N.~Ohta, \textit{Black Holes in Pure
Lovelock Gravities},\ Phys.\ Rev.\ D \textbf{74} (2006) 064001.
[hep-th/0604088].

\bibitem{Dadhich:2012ma} N.~Dadhich, J.~M.~Pons and K.~Prabhu, \textit{On
the static Lovelock black holes},\ Gen.\ Rel.\ Grav.\ \textbf{45} (2013)
1131. arXiv:1201.4994 [gr-qc].

\bibitem{Dadhich:2015ivt} N.~Dadhich, R.~Durka, N.~Merino and O.~Miskovic,
\textit{Dynamical structure of Pure Lovelock gravity}, Phys.\ Rev.\ D
\textbf{93} (2016) 064009. arXiv:1511.02541 [hep-th].

\bibitem{CDIMR} P.K. Concha, R. Durka, C. Inostroza, N. Merino, E.K. Rodr%
\'{\i}guez, \textit{Pure Lovelock gravity and Chern-Simons theory}, Phys.
Rev. D \textbf{94} (2016) 024055. arXiv:1603.09424 [hep-th].

\bibitem{CMR} P.K. Concha, N. Merino, E.K. Rodr\'{\i}guez, \textit{Lovelock
gravity from Born-Infeld gravity theory}, arXiv:1606.07083 [hep-th].

\bibitem{CR2} {P.K. Concha, E.K. Rodr\'{\i}guez, \textit{N=1 supergravity
and Maxwell superalgebras}, JHEP \textbf{1409} (2014) 090. arXiv:1407.4635
[hep-th].}

\bibitem{IW} E. In\"{o}n\"{u}, E.P. Wigner, \textit{On the Contraction of
Groups and Their Representations}, Proc. Nat. Acad. Sci USA \textbf{39}
(1953) 510.

\bibitem{WW} E. Weimar-Woods, \textit{Contractions, Generalized In\"{o}n\"{u}%
-Wigner contractions and deformations of finite-dimensional Lie algebras},
Rev. Mod. Phys. \textbf{12} (2000) 1505.

\bibitem{AI} J.A. de Azc\'{a}rraga, J.M. Izquierdo, \textit{D=3 (p,q)-Poincar%
\'{e} supergravities from Lie algebra expansions}, Nucl. Phys. B \textbf{854
}(2012) 276. arXiv:1107.2569 [hep-th].

\bibitem{KTM} S. Kuzenko, G. Tartaglino-Mazzucchelli, \textit{%
Three-dimensional N=2 (AdS) supergravity and associated supercurrents}, JHEP
\textbf{1112} (2011) 052. arXiv:1109.0496 [hep-th].

\bibitem{KLTM} S. Kuzenko, U. Lindstr\"{o}m, G. Tartaglino-Mazzucchelli,
\textit{Three-dimensional (p,q) AdS superspaces and matter couplings}, JHEP
\textbf{1208} (2012) 024. arXiv:1205.4622 [hep-th].

\bibitem{HS} M. Hatsuda, M. Sakaguchi, \textit{Wess-Zumino term for the AdS
superstring and generalized Inonu-Wigner contraction}, Prog. Theor. Phys.
\textbf{109} (2003) 853. [hep-th/0106114].

\bibitem{AIPV} J.A. de Azc\'{a}rraga, J.M. Izquierdo, M. Pic\'{o}n, O.
Varela, \textit{Generating Lie and gauge free differential (super)algebras
by expanding Maurer-Cartan forms and Chern-Simons supergravity}, Nucl. Phys.
B \textbf{662} (2003) 185 [hep-th/0212347].

\bibitem{AIPV2} J.A. de Azc\'{a}rraga, J.M. Izquierdo, M. Pic\'{o}n, O.
Varela, \textit{Extensions, expansions, Lie algebra cohomology and enlarged
superspaces}, Class. Quant. Grav \textbf{21} (2004) S1375. [hep-th/0401033].

\bibitem{AIPV3} J.A. de Azc\'{a}rraga, J.M. Izquierdo, M. Pic\'{o}n, O.
Varela, \textit{Expansions of algebras and superalgebras and some
applications}, Int. J. Theor. Phys. \textbf{46} (2007) 2738.
[hep-th/0703017].

\bibitem{Sexp} {F. Izaurieta, E. Rodr\'{\i}guez, P. Salgado, \textit{%
Expanding Lie (super)algebras through Abelian semigroups}, J. Math. Phys.
\textbf{47} (2006) 123512. [hep-th/0606215].}

\bibitem{Caroca2010a} R. Caroca, N. Merino, P. Salgado, \textit{S-Expansion
of Higher-Order Lie Algebras}, J. Math. Phys. \textbf{50 }(2009) 013503.
arXiv:1004.5213 [math-ph].

\bibitem{Caroca2010b} R. Caroca, N. Merino, A. Perez, P. Salgado, \textit{%
Generating Higher-Order Lie Algebras by Expanding Maurer Cartan Forms}, J.
Math. Phys. \textbf{50} (2009) 123527. arXiv:1004.5503 [hep-th].

\bibitem{Caroca2011} R. Caroca, N. Merino, P. Salgado, O. Valdivia, \textit{%
Generating infinite-dimensional algebras from loop algebras by expanding
Maurer-Cartan forms}, J. Math. Phys. \textbf{52} (2011) 043519.
arXiv:1311.2623 [math-ph].

\bibitem{CKMN} R. Caroca, I. Kondrashuk, N. Merino, F. Nadal, \textit{%
Bianchi spaces and their three-dimensional isometries as S-expansions of
two-dimensional isometries}, J. Phys. A\textbf{46} (2013) 225201.
arXiv:1104.3541 [math-ph].

\bibitem{AMNT} {L. Andrianopoli, N. Merino, F. Nadal, M. Trigiante, \textit{%
General properties of the expansion methods of Lie algebras}, J. Phys. A
\textbf{46} (2013) 365204. arXiv:1308.4832 [gr-qc].}

\bibitem{ACCSP} M. Artebani, R. Caroca. M.C. Ipinza, D.M. Pe\~{n}afiel, P.
Salgado, \textit{Geometrical aspects of the Lie Algebra S-Expansion Procedure%
}, J. Math. Phys. \textbf{57} (2016) 023516. arXiv:1602.04525 [math-ph].

\bibitem{Durka} R. Durka, \textit{Resonant algebras and gravity},
arXiv:1605.00059 [hep-th].

\bibitem{ILPR} M.C. Ipinza, F. Lingua, D.M. Pe\~{n}afiel, L. Ravera, \textit{%
An Analytic Method for S-expansion involving Resonance and Reduction},
arXiv:1609.05042 [hep-th].

\bibitem{SS} {P. Salgado, S. Salgado, $\mathfrak{so}\left( D-1,1\right)
\otimes \mathfrak{so}\left( D-1,2\right) $\textit{\ algebras and gravity},
Phys. Lett. B \textbf{728} (2013) 5.}

\bibitem{Sorokas1} D.V. Soroka, V.A. Soroka, \textit{Tensor extension of the
Poincar\'{e} algebra}, Phys. Lett. B \textbf{607} (2005) 302.
[hep-th/0410012].

\bibitem{Sorokas2} D.V. Soroka, V.A. Soroka, \textit{Semi-simple extension
of the (super)Poincar\'{e} algebra}, Adv. High Energy Phys. \textbf{2009}
(2009) 234147. [hep-th/0605251].

\bibitem{Sorokas3} D.V. Soroka, V.A. Soroka, \textit{Semi-simple
o(N)-extended super-Poincar\'{e} algebra}, arXiv:1004.3194 [hep-th].

\bibitem{DKGS} R. Durka, J. Kowalski-Glikman, M. Szczachor, \textit{Gauged
AdS-Maxwell algebra and gravity}, Mod. Phys. Lett. A \textbf{26} (2011)
2689. arXiv:1107.4728 [hep-th].

\bibitem{BGKL1} S. Bonanos, J. Gomis, K. Kamimura, J. Lukierski, \textit{%
Maxwell Superalgebra and Superparticle in Constant Gauge Backgrounds}, Phys.
Rev. Lett. \textbf{104} (2010) 090401. arXiv:0911.5072 [hep-th].

\bibitem{BGKL2} S. Bonanos, J. Gomis, K. Kamimura, J. Lukierski, \textit{%
Deformations of Maxwell Superalgebras and Their Applications}, J. Math.
Phys. \textbf{51} (2010) 102301. arXiv:1005.3714 [hep-th].

\bibitem{L} J. Lukierski, \textit{Generalized Wigner-Inonu Contractions and
Maxwell (Super)Algebras}, Proc. Stekl. Inst. Math. \textbf{272} (2011) 1-8.
arXiv:1007.3405 [hep-th].

\bibitem{KL} K. Kamimura, J. Lukierski, \textit{Supersymmetrization Schemes
of D=4 Maxwell Algebra}, Phys. Lett. B \textbf{707} (2012) 292.
arXiv:1111.3598 [math-ph].

\bibitem{CRS} P.K. Concha, E.K. Rodr\'{\i}guez, P. Salgado, \textit{%
Generalized supersymmetric cosmological term in N=1 supergravity}, JHEP
\textbf{08} (2015) 009. arXiv:1504.01898 [hep-th].

\bibitem{AKL} {J.A. de Azcarraga, K. Kamimura, J. Lukierski, \textit{%
Generalized cosmological term from Maxwell symmetries}, Phys. Rev. D \textbf{%
83} (2011) 124036. arXiv:1012.4402 [hep-th].}

\bibitem{CIRR} P.K. Concha, M.C. Ipinza, L. Ravera, E.K. Rodr\'{\i}guez,
\textit{On the Supersymmetric Extension of Gauss-Bonnet like Gravity}, JHEP
(2016). arXiv:1607.00373 [hep-th].

\bibitem{DFIMRSV} J. Diaz, O. Fierro, F. Izaurieta, N. Merino, E. Rodr\'{\i}%
guez, P. Salgado, O. Valdivia, \textit{A generalized action for
(2+1)-dimensional Chern-Simons gravity}, J. Phys. A \textbf{45} (2012)
255207. arXiv:1311.2215 [gr-gc].

\bibitem{HR} S. Hoseinzadeh, A. Rezaei-Aghdam, (\textit{2+1)-dimensional
gravity from Maxwell and semi-simple extension of the Poincar\'{e} gauge
symmetric models}, Phys. Rev. D \textbf{90 }(2014) 084008. arXiv:1402.0320
[hep-th].

\bibitem{AF} R. D'Auria, P. Fr\'{e}, \textit{Geometric Supergravity in d=11
and Its Hidden Supergroup}, Nucl. Phys. B \textbf{201} (1982) 101.

\bibitem{Green} M.B. Green, \textit{Supertranslations, Superstrings and
Chern-Simons Forms}, Phys. Lett. B \textbf{223} (1989) 157.

\bibitem{AILW} {J.A. de Azcarraga, J.M. Izquierdo, J. Lukierski, M.
Woronowicz, \textit{Generalizations of Maxwell (super)algebras by the
expansion method}, Nucl. Phys. B \textbf{869} (2013) 303. arXiv:1210.1117
[hep-th]. }

\bibitem{CR1} {P.K. Concha, E.K. Rodr\'{\i}guez, \textit{Maxwell
Superalgebras and Abelian Semigroup Expansion}, Nucl. Phys. B \textbf{886}
(2014) 1128. arXiv:1405.1334 [hep-th]. }

\bibitem{CDMR} P.K. Concha, R. Durka, N. Merino, E.K. Rodr\'{\i}guez,
\textit{New family of Maxwell like algebras}, Phys. Lett. B \textbf{759}
(2016) 507. arXiv:1601.06443 [hep-th].

\bibitem{SSV} P. Salgado, R. J. Szabo, O. Valdivia, \textit{Topological
gravity and transgression holography}, Phys. Rev. D \textbf{89 }(2014)
084077. arXiv:1401.3653 [hep-th].

\bibitem{BC1} M. Blagojevic, B. Cbetkovic, \textit{Black hole entropy in 3D
gravity with torsion}, Class. Quant. Grav. \textbf{23} (2006) 4781.
[gr-qc/0601006].

\bibitem{BC2} M. Blagojevic, B. Cbetkovic, \textit{Covariant description of
the black hole entropy in 3D gravity}, Class. Quant. Grav. \textbf{24}
(2007) 129. [gr-qc/0607026].

\bibitem{TZ} P.K. Townsend, B. Zhang, \textit{Thermodynamics of "exotic" Ba%
\~{n}ados-Teitelboim-Zanelli black holes}, Phys. Rev. Lett. \textbf{110}
(2013) 241302. arXiv:1302.3874 [hep-th].
\end{thebibliography}
\end{document}